\documentclass[preprint,12pt]{elsarticle}

\usepackage{amssymb}

\usepackage{graphicx,amsmath}
\usepackage[margin=34mm]{geometry}
\usepackage{longtable}
\usepackage{booktabs}
\usepackage[hyphens]{url}
\usepackage[colorlinks=true,linkcolor=black, citecolor=blue, urlcolor=blue]{hyperref}

\usepackage{soul}
\usepackage{tabularx}
\usepackage{ltablex}
\usepackage{xcolor}
\usepackage{braket}
\usepackage{footnote}
\usepackage{xr}

\usepackage{dcolumn}
\usepackage{bm}
\usepackage{multirow}

\biboptions{square,sort,compress}

\usepackage{lineno}

\begin{document}

\begin{frontmatter}


\title{The long-term impact of ranking algorithms in growing networks}

\author[label1]{Shilun Zhang}
 
\author[label1,label2,label3]{Mat{\'u}{\v{s}} Medo}
  
\author[label1,label4]{Linyuan L\"u}
\ead{linyuan.lv@uestc.edu.cn}

\author[label1,label5]{Manuel Sebastian Mariani}
\ead{manuel.mariani@business.uzh.ch}

\address[label1]{Institute of Fundamental and Frontier Sciences, University of Electronic Science and Technology of China, 610051 Chengdu, PR China}
\address[label2]{Department of Radiation Oncology, Inselspital, Bern University Hospital and University of Bern, 3010 Bern, Switzerland}
\address[label3]{Department of Physics, University of Fribourg, 1700 Fribourg, Switzerland}
\address[label4]{Alibaba Research Center for Complexity Sciences, Hangzhou Normal University, 311121 Hangzhou, PR China}
\address[label5]{URPP Social Networks, Universit\"at Z\"urich, 8050 Z\"urich, Switzerland}


 \begin{abstract}
When we search online for content, we are constantly exposed to rankings. For example, web search results are presented as a ranking, and online bookstores often show us lists of best-selling books. While popularity-based ranking algorithms (like Google's PageRank) have been extensively studied in previous works, we still lack a clear understanding of their potential systemic consequences. In this work, we fill this gap by introducing a new model of network growth that allows us to compare the properties of the networks generated under the influence of different ranking algorithms. We show that by correcting for the omnipresent age bias of popularity-based ranking algorithms, the resulting networks exhibit a significantly larger agreement between the nodes' inherent quality and their long-term popularity, and a less concentrated popularity distribution. To further promote popularity diversity, we introduce and validate a perturbation of the original rankings where a small number of randomly-selected nodes are promoted to the top of the ranking. Our findings move the first steps toward a model-based understanding of the long-term impact of popularity-based ranking algorithms, and could be used as an informative tool for the design of improved information filtering tools.
\end{abstract}

\begin{keyword}
Complex networks \sep Ranking \sep Popularity and quality \sep Popularity inequality \sep Algorithmic bias
\end{keyword}

\end{frontmatter}


\section{Introduction}

Ranking algorithms allow us to efficiently sort the massive 
amount of online information, aiming to quickly provide us with the relevant items that fit our needs.
Due to the ubiquity of ranking, the implications of ranking-based information 
filtering tools such as search engines~\cite{brin1998anatomy} and recommendation 
systems~\cite{lu2012recommender} for our society are widely
debated~\cite{hindman2003googlearchy,cho2004impact,diaz2008through,bozdag2013bias}. 
Even when free from apparent manipulation, the rankings that we
are exposed to in online platforms and information 
repositories are influenced by social processes and, at 
the same time, influence social processes themselves~\cite{scholtes2014social}.
To provide a few examples, rankings can heavily impact on the
eventual popularity of movies and songs in 
cultural markets~\cite{salganik2006experimental}, affect the 
attention received by products in online e-commerce platforms~\cite{jannach2009case, sharma2015estimating,zeng2015modeling}, increase the sales of top-ranked dishes in restaurants~\cite{cai2009observational},
and even influence the choices of undecided electors and, as a result, the outcome 
of political elections~\cite{epstein2015search}. 
Understanding the potential systemic impact of ranking algorithms and correcting their potential flaws becomes 
therefore a critical issue in diverse contexts.

A robust finding in previous experiments of diverse nature~\cite{lawrence1999accessibility,cho2004impact,fortunato2006topical,salganik2006experimental,cai2009observational,salganik2009web} is 
that the current ranking position of an item (or, generally, its current popularity) heavily influences its eventual
popularity or success.
As a consequence, one of the key challenges is to assess whether in a given system,
the final popularity of an item is a reliable proxy for
its ``quality" or ``fitness", where quality is interpreted as the success the item would have in absence of social influence mechanisms~\cite{salganik2006experimental}.
In other words, it becomes critical to determine whether the adoption of a given ranking algorithm by a given system
allows high-quality nodes to experience larger success than low-quality nodes. If this is not the case, we may conclude that the adopted ranking algorithm has a negative impact on the system, as it may prevent high-quality nodes from becoming popular and allow low-quality nodes to stand out.

Cho~\cite{cho2004impact} pointed out 
that ranking algorithms and search engines that favor already popular items
create a strong ``popularity bias"~\cite{fortunato2006topical} -- also dubbed as ``search-engine bias"~\cite{cho2004impact}, and 
``googlearchy"~\cite{hindman2003googlearchy} -- such that only the already popular nodes can receive
substantial attention in the next future, whereas recent high-quality nodes go essentially 
unnoticed~\cite{cho2004impact}. This bias can amplify initial differences between the items' popularity,
leading some items to a disproportionately high popularity regardless of their
quality. 
O'Madadhain et al.~\cite{o2005eventrank} emphasized that static ranking algorithms (like Google's PageRank~\cite{brin1998anatomy}) are based on time-aggregate network representations and, for this reason, they do not respect the sequence of events that led to the formation of the network. To solve this issue, they introduced~\cite{o2005eventrank} and validated~\cite{o2005prediction} a network-based ranking algorithm, called EventRank, that takes into account the detailed sequence of events, in a similar spirit as the recent literature on temporal networks~\cite{scholtes2014causality,scholtes2016higher,xu2016representing,liao2017ranking}.
Salganik et al.~\cite{salganik2006experimental,salganik2009web} found that in artificial cultural markets, showing the items' ranking by popularity to consumers significantly impacts on the items' final popularity.
Recent studies~\cite{weng2012competition, qiu2017limited} emphasized the individuals' limited attention as a determinant for the viral popularity of low-quality items.
Other recent works focused on modelling the interplay between quality/talent and popularity for diverse types of agents, including websites~\cite{kong2008experience}, scientific papers~\citep{medo2011temporal,wang2013quantifying}, researchers~\cite{medo2016model,sinatra2016quantifying}, and  bestseller books~\cite{yucesoy2018success}.
Both model-based~\cite{kong2008experience,medo2011temporal} and 
experimental results~\cite{salganik2006experimental} indicate that in presence of social influence, the relation between popularity and quality is highly non-linear, meaning that a small variation of quality leads to large variations in popularity.

While both the relation between popularity and quality and the 
search-engine bias in the Web have attracted considerable attention from previous research,
the abilities of different ranking algorithms to promote quality in a given system are typically not compared with each other.
The main reason is that the intrinsic node quality is typically inaccessible in the real world, and we lack a clear understanding of how the interplay between ranking and quality shapes the growth of a given system.
As a result, we still lack a general framework to assess the long-term impact of different ranking algorithms.

The main goal of this paper is to move the first steps toward filling this gap.
To this end, building on existing models of network growth~\cite{fortunato2006scale} and 
popularity dynamics~\cite{ciampaglia2018algorithmic}, we introduce a growing directed-network model
where each node, when choosing the nodes to point to, is driven either by the results of a given adopted ranking algorithm or by quality. The nodes' sensitivity to quality is a homogeneous parameter of the model~\cite{ciampaglia2018algorithmic}.
Crucially, different adopted ranking algorithms lead to different properties of the final network. 
We use the model to address the following questions: Will
a given algorithm facilitate or impede the success of high-quality nodes in the system? Is a given algorithm useful in discovering high-quality nodes? Does it lead to uneven, highly-concentrated popularity distribution?

We postulate that a good ranking algorithm should lead to a
network where: (1) node long-term popularity strongly correlates with their quality (\emph{quality promotion}); (2) nodes' score strongly correlates with their quality (\emph{quality detection}); (3) the node popularity distribution is relatively diverse (\emph{diversity promotion}).
The quality promotion and detection properties favor the algorithms that help the system to improve the popularity-quality correlation (quality promotion)~\cite{ciampaglia2018algorithmic} and help the nodes to find high-quality nodes (quality detection)~\cite{mariani2015ranking}. 
The diversity promotion favors the algorithms that distribute popularity more evenly across the nodes, making it easier for the nodes to find quality nodes that are not among the most popular ones~\cite{zhou2010solving}. 

In fact, the synthetic networks generated with our ranking-based growth model can be interpreted as benchmark graphs for ranking algorithms. Our focus on quality promotion, quality detection, and diversity promotion makes our validation framework for ranking algorithm fundamentally different from existing benchmarking techniques which focus on the ability of the algorithms to identify structurally vital nodes~\cite{morone2015influence}, find those nodes that maximize the reach of a spreading process~\cite{lu2016vital,radicchi2016leveraging}, single out expert-selected important nodes~\cite{mariani2016identification,mariani2018early}, or respect sets of axioms~\cite{o2005prediction,boldi2014axioms,schoch2016re,boldi2017rank}.

We find that in networks that adopt cumulative popularity (as measured by the number of incoming links -- node indegree~\cite{newman2010networks}) as the ranking algorithm, the correlation between node popularity and quality strongly depends not only on the nodes'
sensitivity to quality, but also on their willingness to select low-ranked nodes (``exploration cost'' in~\cite{ciampaglia2018algorithmic}).
A popularity metric that is not biased by node age~\cite{newman2009first,mariani2016identification} (called rescaled indegree in~\cite{mariani2016identification}) leads to networks where both the final nodes' popularity and
the node score are
significantly better correlated with node quality, and the final popularity
distribution is significantly more diverse.  Interestingly, when the exploration cost is large, networks
that adopted a random ranking of the nodes exhibit even higher indegree-quality correlation than networks that adopted a popularity-based 
ranking: while popularity-based ranking algorithms are always useful for the nodes to discover high-quality content, 
they may accelerate the dissemination of low-quality content when individuals rely too heavily on them.  

To further promote popularity diversity, we introduce and validate a ranking algorithm -- the ranking by rescaled indegree with random promotion -- where the original ranking by rescaled indegree is ``perturbed" by promoting a small number of randomly-selected nodes to the top-$10$ or the top-$20$ of the ranking. Such a perturbation has a deterministic component (the number of nodes that are promoted is fixed) and a noisy one (the promoted nodes are chosen at random), and it allows us to study the impact of a small amount of noise on systemic properties, in a similar spirit as previous studies that investigated the impact of noise on democratic consensus promotion in animal groups~\cite{couzin2011uninformed}, dynamical influence detection~\cite{jiang2016directed}, and the performance of human groups in coordination problems~\cite{shirado2017locally}.
We find that with respect to the ranking by rescaled indegree, the ranking by rescaled indegree with random promotion generates networks with increased popularity diversity. Intriguingly, we find that the random promotion can have a marginal or a negative impact on the agreement between nodes' final popularity and quality, depending on the model parameters.

This work provides the first systematic comparison of network-based ranking algorithms with respect to their long-term systemic effects. It complements the question of whether a given ranking algorithm is able to detect important nodes~\cite{lu2016vital,liao2017ranking} with the question of whether high-quality nodes will stand out in a system that adopted that given metric. 
It reveals that suppressing the bias by node age of popularity-based metrics is beneficial to quality promotion, detection, and popularity diversity.

The manuscript is organized as follows. Section~\ref{sec:model_algorithms} introduces our model of network growth together with the ranking algorithms considered here, and the ranking evaluation criteria. Section~\ref{sec:results} presents the results of our numerical simulations both for the basic model (Sections~\ref{sec:qp}-\ref{sec:d}) and for a variant of the model with node removal (Section~\ref{sec:r}), together with an application of our model to a real information network of scientific papers (Section~\ref{sec:aps}). Section~\ref{sec:discussion} is dedicated to a discussion of our results and their implications for algorithmic evaluation and the quality-popularity relation in information systems. Appendices~A-B conclude the main text. The Supplementary Material (SM) file is available online -- figures whose label contains an ``S" (e.g., Fig.~S1) can be found in the SM file.

\section{Model and ranking algorithms}
\label{sec:model_algorithms}

In this Section, we introduce the model of network growth (Section~\ref{sec:model}), the ranking algorithms considered in this paper (Section~\ref{sec:algo}), and the metrics used to evaluate the long-term impact of ranking algorithms (Section~\ref{sec:metrics}).

\subsection{The model of network growth}
\label{sec:model}

We focus here on monopartite directed networks. 
Our model is meant to represent a social or information network where the number of nodes grows with time.
In the model, each node $i$ is endowed with a quality parameter $q_i$ which quantifies its
attractiveness to new incoming connections in absence of ranking influence. Before generating 
each network, we choose the ranking algorithm $\mathcal{A}$ that influences the growth -- equivalently, as 
the nodes choose their links\footnote{The network's directed links might be interpreted as friendship or follower relationships in online social networks, or as citations between documents in information networks.} based on the node ranking by $\mathcal{A}$, we say that the system ``has adopted" algorithm $\mathcal{A}$.
We introduce a model which features three essential elements: 
\begin{enumerate}
\item \emph{Growth.} At each time step, one new node enters the system. Nodes can thus be labeled directly by the time step in which they appeared. Each new node creates $m$ directed links to $m$ different preexisting nodes\footnote{Self-loops and multiple links between a given pair of nodes are prohibited.}.
 \item \emph{Ranking-driven attachment.} With probability $\beta$, node $t$ chooses its target according to the probability
\begin{eqnarray}\label{equ:P_rank}
P^{(\mathcal{A})}(j,t)=\frac{r_{j}(t)^{-\alpha}}{\sum_{l=1}^{t-1} r_{l}(t)^{-\alpha}},
\label{ranking}
\end{eqnarray}
where $r_{j}(t)$ is the ranking position of node $j$ at time $t$ according to $\mathcal{A}$; the real number $\alpha\geq 0$ is a tunable model parameter referred to as \emph{exploration cost} by Ciampaglia et al.\cite{ciampaglia2018algorithmic}: Large values of $\alpha$ imply that the nodes are only willing to connect to the top-nodes by the adopted ranking algorithm, whereas low values of $\alpha$ allow the nodes to also connect to low-ranked nodes.
 \item \emph{Quality-driven attachment.} With probability $1-\beta$, node $t$ chooses its target according to the probability
\begin{eqnarray}\label{equ:P_quality}
P^{(q)}(j,t)=\frac{q_{j}}{\sum_{l=1}^{t-1}q_l}.
\label{quality}
\end{eqnarray}
\end{enumerate}
Our model reduces to the model by Fortunato et al.~\cite{fortunato2006scale} in the special case $\beta=1$; node
quality and quality-driven attachment are novel elements with respect
to Fortunato et al.'s model~\cite{fortunato2006scale}. 
Differently from the recent popularity dynamics model by Ciampaglia 
et al.~\cite{ciampaglia2018algorithmic} which considers a cultural 
market composed of a fixed number $N$ of items, our model represents 
a network that grows with time. Differently from previous
works~\cite{fortunato2006topical,medo2011temporal,ciampaglia2018algorithmic}, we aim to use our 
growing network model to compare the long-term properties of the networks generated based on different ranking algorithms.

\subsection{Ranking algorithms}
\label{sec:algo}

As our main goal is to uncover the long-term implications of the temporal bias 
of static centrality metrics and the benefits from suppressing such bias, we focus here on indegree and rescaled indegree~\cite{mariani2016identification}. Besides, we introduce a random promotion mechanism which ``perturbes" the original ranking by rescaled indegree by promoting randomly-selected nodes to the top of the ranking, and we consider a random ranking of the node as a baseline.
We provide below the details of these four ranking algorithms. 
\begin{enumerate}
\item \emph{Ranking by indegree, $k$.} The indegree\footnote{In the following, we will use 
interchangeably ``indegree", ``popularity", and ``cumulative popularity". This is because,
in our simple setting, the incoming links received by a node are the only available information on its ``popularity". The situation
might be different in a real online system where, for example, the popularity of a video can be quantified by the
number of downloads, by the number of views, by the number of shares, etc.} of a node is defined as the number of 
incoming connections received by that node~\cite{newman2010networks}. We simply rank the nodes in order of 
decreasing indegree $k$, which is arguably the simplest way to rank the nodes in a directed network~\cite{newman2010networks}.
In growing networks, node indegree is strongly biased by node age~\cite{newman2009first,mariani2016identification,liao2017ranking}, as 
confirmed by numerical simulations and analytic computations with our model (see \ref{sec:relation}). 
\item \emph{Ranking by (age-)rescaled indegree, $R(k)$.} We rank the nodes in order of decreasing age-rescaled 
indegree~\cite{mariani2016identification} $R(k)$. The rescaled indegree is built on indegree by 
requiring that node score is not biased by node age. More specifically, for each node $i$, we 
consider a reference set $\mathcal{R}_i:=\{i-\Delta/2,\dots,i+\Delta/2\}$ of $\Delta+1$ nodes of similar age as node $i$ -- we set $\Delta=0.01\,N$. We 
compute the mean $\mu_i(k)$ and the standard deviation $\sigma_i(k)$ of node indegree within this 
reference set. In formulas, 
\[
\begin{split}
\mu_i(k)&=\frac{1}{\Delta+1}\,\sum_{j\in\mathcal{R}_i}k_j, \\
\sigma_i(k)&=\sqrt{\frac{1}{\Delta+1}\,\sum_{j\in\mathcal{R}_i}(k_j-\mu_i(k))^2}
\end{split}
\].
The rescaled indegree score $R_i(k)$ of node $i$ is given by the $z$-score~\cite{mariani2016identification}
\begin{equation}
R_i(k)=\frac{k_i - \mu_i(k) }{\sigma_i(k)}.
\end{equation}
The rescaled indegree of a given node thus quantifies how larger the node's indegree is
with respect to nodes of similar age, in units of standard deviations. 
Using the $z$-score to normalize static metrics of node importance is customary in scientometrics~\cite{lundberg2007lifting,zhang2014comparison,vaccario2017quantifying} where scholars aim to gauge the impact of a given scientific paper independently of its field and publication date~\cite{waltman2016review}.
Besides, in citation networks, the rescaled indegree
allows us to identify significantly earlier important papers~\cite{newman2009first,mariani2016identification}, movies~\cite{ren2018randomizing} and patents~\cite{mariani2018early} with respect to citation count.
\item \emph{Ranking by age-rescaled indegree with Random Promotion, ($R(k)$+RP).} While age-rescaled indegree substantially suppresses the cumulative advantage of older nodes~\cite{newman2009first,mariani2016identification}, it may still amplify the advantage of some nodes that received quickly many connections with respect to nodes of similar age. To reduce this potential problem and further increase diversity, we introduce the ranking by rescaled indegree with random promotion ($R(k)$+RP): We rank the nodes by age-rescaled indegree, $R(k)$, and we ``promote'' $P$ randomly-selected nodes to the top-$T$ of the ranking. Therefore, a fraction $\eta:=P/T$ of nodes in the top-$T$ by the ranking is purely determined by noise. 
By placing some previously overlooked nodes at the top of the ranking, this mechanism gives these nodes enhanced visibility and, therefore, an additional opportunity to attract some links.
In the following, we show results for $T=10, \eta=0.5$; results for other values of $T$ and $\eta$ ($\eta=P/T=5/10, 8/10, 10/10, 10/20, 16/20, 20/20$) are shown in the Supplementary Material (Supplementary Figures S11--S13 and S19-S24).
\item \emph{Random ranking.} The nodes are ranked at random. The resulting ranking is used as a baseline to understand for which parameter values the final indegree-quality correlation benefits from the rankings by indegree and rescaled indegree.
\end{enumerate}

For all the ranking algorithms considered above, if two or more nodes happen to have the same value of the score, their relative order is determined at random.
Choosing the quality distribution deserves some attention. 
A simple mean-field approximation shows that for $\beta=0$, the final
popularity of the nodes $k$ is expected to be proportional to node quality; simulation results show that for $\beta>0$, node final popularity is a power-law function of node quality (see \ref{sec:relation} for details). 
Motivated by this property, to mimic the broad popularity distributions typically observed in real data~\cite{newman2010networks}, we choose a Pareto distribution of the quality 
values $q$ (see \ref{sec:details} for details). This choice leads indeed to broad indegree distributions as 
shown in Fig.~S1. 
We refer to \ref{sec:details} for all the simulation details.

We emphasize that the rescaling procedure described above is only one among the possible ways to design a time-aware ranking algorithm~\cite{liao2017ranking}. Already in 2005, O'Madadhain and Smyth ~\cite{o2005eventrank} recognized that static centrality metrics are based on time-aggregate network representations and, for this reason, they do not respect the sequence of events that led to the formation of the network. To solve this issue, O'Madadhain~\textit{et al.} introduced~\cite{o2005eventrank} and validated~\cite{o2005prediction} a ranking algorithm based on the detailed contact time-series, acting as precursors of the recent stream of literature on centrality in temporal networks based on time-preserving paths~\cite{scholtes2014causality,scholtes2016higher,xu2016representing,liao2017ranking}. 
We refer to~\cite{liao2017ranking} for a review of time-dependent ranking algorithms in complex networks. Nevertheless, we focus on the age-rescaled indegree here because of its simplicity and effectiveness in suppressing the indegree's bias towards old nodes~\cite{mariani2016identification,liao2017ranking}. Testing alternative time-dependent ranking algorithms within our model-generated benchmark graphs is an interesting possibility for future research.

\subsection{Evaluating the algorithms' long-term impact}
\label{sec:metrics}

To assess the long-term impact of different algorithms, we grow random networks based on the model described above for each ranking algorithm $\mathcal{A}$. The algorithm determines, at any time, the ranking of the nodes
that, in turn, determines the probability that a node receives a new connection, according to Eq. \eqref{ranking}.
We refer to the networks generated with the algorithm $\mathcal{A}$ as to the \emph{$\mathcal{A}$-generated networks}. 
Ideally, we would expect a good ranking algorithm $\mathcal{A}$ to exhibit the three main properties introduced 
above: (i) \emph{Quality promotion}: The algorithm generates networks where the final popularity of the nodes
strongly correlates with their quality; (ii) \emph{Quality detection}: The algorithm is effective in
identifying high-quality nodes; (iii) \emph{Popularity diversity}: The algorithm generates networks where the popularity
is not strongly concentrated among few nodes. 
In the following, we introduce three classes of observables to quantify these three properties. 

\begin{figure*}[t]
\centering
\includegraphics[scale=0.4]{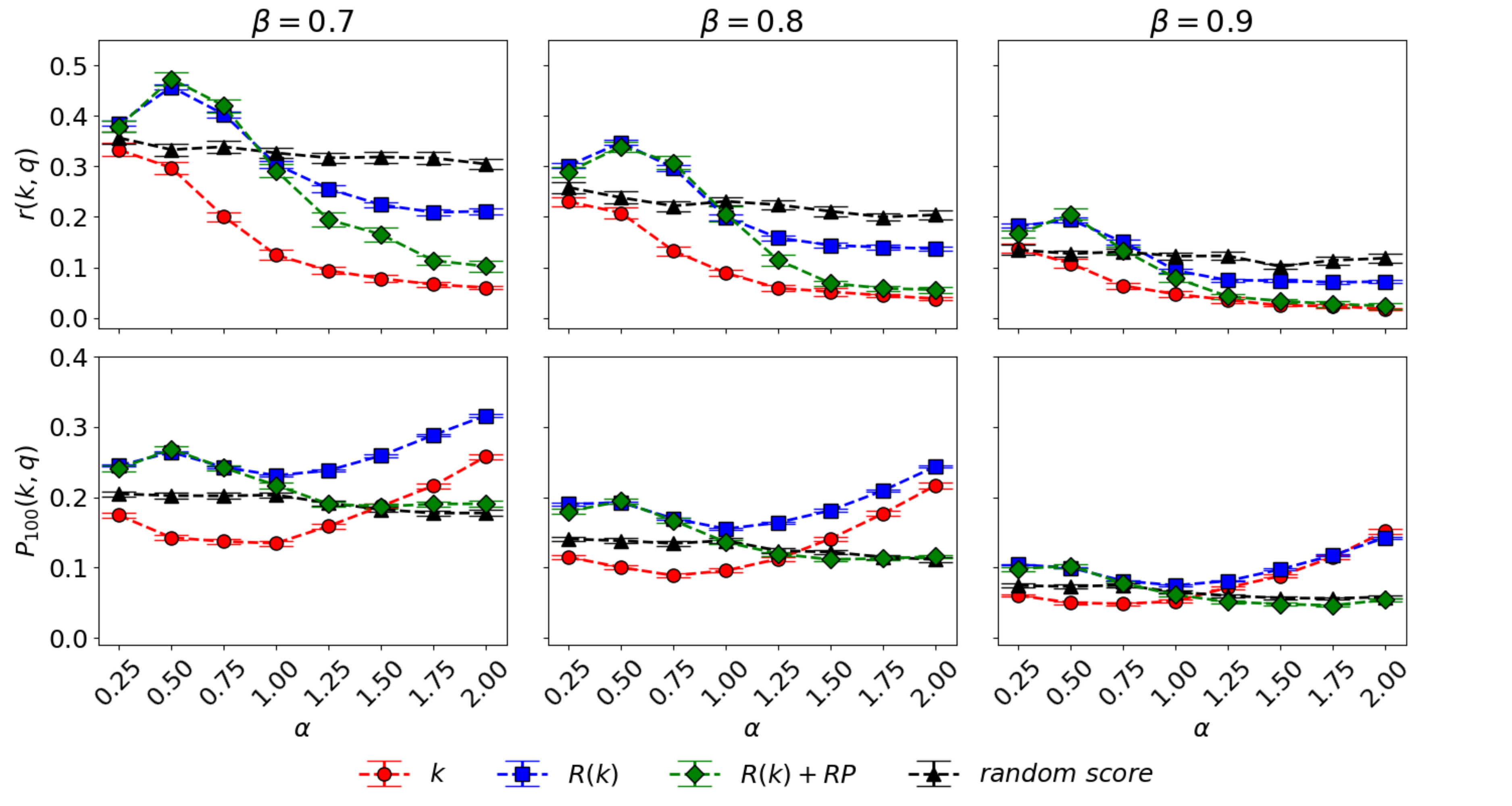}
\caption{Quality promotion as measured by $r(k, q)$ (the Pearson's linear correlation between node indegree $k$ and node quality $q$ -- \emph{top} panels), and $P_{100}(k, q)$ (the precision of node indegree $k$ in identifying the top-$100$ nodes by quality $q$ -- \emph{bottom} panels): comparison between indegree-generated (red circles), $R(k)$-generated (blue squares), $R(k)+RP$-generated (green rhombuses), and random-generated networks (triangles). The three columns correspond, from left to right, to $\beta=0.7,0.8,0.9$, respectively. Results are averaged over $100$ realizations; the error bars represent the standard error of the mean. 
}
\label{kq_corr2}
\end{figure*}

\paragraph{Quality promotion}
For a given ranking algorithm $\mathcal{A}$, we evaluate how well the final popularity $k$ of the nodes 
reproduces the inherent quality values $q$ for $\mathcal{A}$-generated networks.
We calculate the Pearson's linear correlation $r^{\mathcal{A}}(k,q)$ between node popularity $k$ and node quality $q$.
While this metric takes all the nodes into account, we are also interested in the algorithm's ability to promote the top-quality nodes.
To this end, we measure the precision $P_{100}^{\mathcal{A}}(k,q)$ defined as the fraction of 
nodes that are placed in the top-$100$ of both the ranking by $k$ and the ranking by $q$. Nevertheless, quality 
promotion is not sufficient alone to evaluate the metrics because, by construction, the nodes have a non-zero 
probability to choose their targets based on quality. As a consequence, even the random ranking of the
nodes produces networks with non-zero indegree-quality correlation. Such correlation increases as the nodes' sensitivity 
to quality increases (i.e., as $\beta$ decreases) -- see Fig.~\ref{kq_corr2} and the related discussion below. 
However, the random ranking is useless to find valuable nodes in the system. 
For this reason, we study not only quality promotion, but also quality detection.

\paragraph{Quality detection}
For a given ranking algorithm $\mathcal{A}$, we evaluate how well the scores by $\mathcal{A}$ reproduce 
the inherent quality values $q$ for $\mathcal{A}$-generated networks. 
To this end, we measure the Pearson's linear correlation $r^{\mathcal{A}}(s,q)$ between the node-level 
scores $s$ produced by the algorithm $\mathcal{A}$ (measured at the end of the network growth) and
node quality $q$ for $\mathcal{A}$-generated networks. In parallel, we also measure the precision~\cite{lu2012recommender} $P_{100}^{\mathcal{A}}(s,q)$ of the algorithm, defined as the fraction of
nodes that are placed in the top-$100$ of both the ranking by $s$ and the ranking by $q$.

\paragraph{Diversity}
To quantify the ability of the algorithms to evenly spread popularity across the network's nodes,
we measure the indegree's Herfindahl index \cite{herfindahl1959copper} $H(\mathbf{k})$.
\begin{eqnarray}
H(\mathbf{k})=\sum_{i=1}^{N}\biggl(\frac{k_i}{L} \biggr)^2,
\end{eqnarray}
where $L$ is the total number of links.
The index is proportional to the variance of the network's indegree distribution: the smaller $H(\mathbf{k})$,
the less concentrated indegree is within a restricted group of nodes. More specifically, the index ranges between $H_{min}=1/N$ (egalitarian
network where all the nodes have indegree equal to $L/N$) and $H_{max}=1$ (network where one node has 
indegree $L$, and all the other nodes have indegree equal to zero). We define $N_{eff}(\mathbf{k})=1/H(\mathbf{k})$ as 
the ``effective number of nodes" that received incoming links. Such number $N_{eff}$ is equal to $1$ if one single
node received all the incoming links, and it is equal to $N$ if all the nodes received the same amount of links. 
We posit that a good ranking algorithm should not produce too concentrated networks, and its generated networks should therefore
exhibit relatively large values of $N_{eff}$.

\section{Results}
\label{sec:results}

We grow networks of $N=10,000$ nodes according to the model described above; we 
refer to \ref{sec:details} for all the simulation details. Here, we show the results for 
node outdegree $m=6$; the results for $m=3$ are commented below and shown in the SM (Figs.~S2--S13).
Our goal is to compare the properties of networks generated by the four ranking algorithms defined in Section~\ref{sec:model_algorithms}, 
according to the observables described above.

\begin{figure*}[t]
\centering
\includegraphics[scale=0.3]{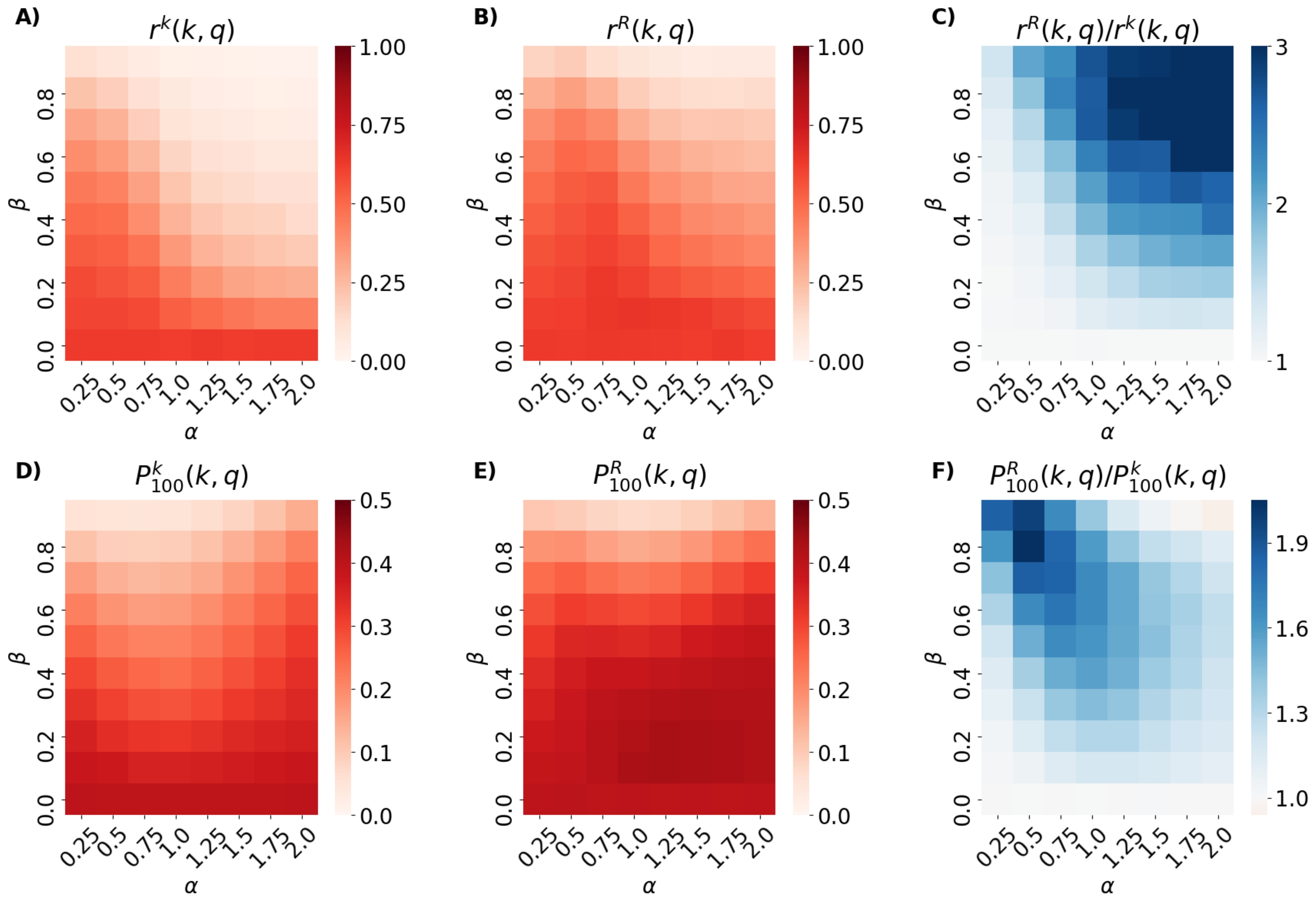}
\caption{Quality promotion as measured by $r(k,q)$ (the Pearson's linear correlation between 
node indegree $k$ and node quality $q$ -- \emph{top} panels), and $P_{100}(k, q)$ (the precision of node indegree $k$ in identifying the top-100 nodes by quality $q$ -- \emph{bottom} panels): comparison between indegree-generated and $R(k)$-generated 
networks. (A-B): $r(k, q)$ for indegree-generated ($r^k(k, q)$, panel A) and $R(k)$-generated ($r^R(k, q)$, panel B) networks,
as a function of the model parameters $\alpha$ (exploration cost) and $\beta$ (reliance on ranking). (C): Ratio $r^{R}(k,q)/r^{k}(k,q)$ as a function of the model parameters. (D-E): $P_{100}(k, q)$ for indegree-generated ($P^k_{100}(k, q)$, panel D) and $R(k)$-generated ($P^R_{100}(k, q)$, panel E) networks, as a function of the model parameters. (F): Ratio $P^R_{100}(k, q)/P^k_{100}(k, q)$ as a function of the model parameters. Results are averaged over $500$ realizations.}
\label{kq_corr}
\end{figure*}

\subsection{Quality promotion: Impact of the age bias suppression}
\label{sec:qp}

Fig. \ref{kq_corr}A shows the indegree-quality Pearson's linear correlation $r(k,q)$ in indegree-generated networks, as a function of the model parameters $\alpha$ and $\beta$. 
The correlation between final popularity and quality is sensitive to both $\alpha$ and $\beta$.
As $\alpha$ grows, it becomes harder for low-ranked high-quality nodes to
acquire new incoming connections, which results in a lower indegree-quality correlation.
The (approximately) monotonous dependence of $r^k(k,q)$ on $\alpha$ was not found for the model of a static market
\footnote{By static market, we mean a collection of a fixed number of items. This is different from our growing network model where, at each time step, a new node enters the system and connects to the preexisting nodes.}
by Ciampaglia et al. \cite{ciampaglia2018algorithmic},
which indicates that it is a consequence of the network's growth.
As $\beta$ grows, the nodes become less sensitive to quality and, as a direct consequence, the indegree-quality correlations deteriorates.

Fig. \ref{kq_corr}B shows the correlation $r^{R}(k,q)$ between node indegree $k$ and node quality in $R(k)$-generated networks. 
The figure shows that by adopting the age-rescaled metric $R(k)$, the indegree-quality correlation stays large for a broader parameter region. For example, we observe values of $r^R(k,q)$ as high\footnote{As opposed to $r^k(k,q)=0.12$ observed for
indegree-generated networks for the same pair of $(\alpha,\beta)$ values.}
as $0.35$ when $(\alpha,\beta)=(1,0.7)$,
which corresponds to a scenario where the nodes are driven by quality only three times out of ten. 

To visually appreciate the parameter regions where the indegree-quality correlations significantly differ among the two classes of networks,
we represent the heatmap of the ratio $r^{R}(k,q)/r^{k}(k,q)$ (Fig. \ref{kq_corr}C).
When $\beta=0$, all the nodes are only sensitive to quality, and the plotted ratio is thus one by definition (on average) because node ranking has no influence.
As soon as $\beta>0$, the nodes become driven both by quality and by ranking, which makes it possible to reveal 
the differences between the networks grown with different algorithms.
We find that rescaled indegree produces networks with a higher indegree-quality correlation for all the parameter
space; we observe the largest advantage of the rescaled indegree in terms of quality promotion
for the region where both $\alpha$ and $\beta$ are relatively large -- i.e., in the region where the nodes are unwilling to choose low-ranked nodes
and, at the same time, are highly sensitive to ranking. 
Analogous heatmaps for the precision metrics (Figs.~\ref{kq_corr}D--F) show that the indegree's precision in
$R(k)$-generated networks is systematically larger than that in indegree-generated networks.
Differently from Fig.~\ref{kq_corr}C, Fig.~\ref{kq_corr}F shows that
the largest gaps between the precision in $R(k)$- and indegree-generated networks occur in the 
small $\alpha$, large $\beta$ region. 
We discuss the reasons behind the different trend for correlation and precision in the next paragraph.

\subsection{Quality promotion: comparing the four ranking algorithms}

Fig.~\ref{kq_corr} indicates that adopting the rescaled indegree allows us to better promote node quality for a broad range of model parameters. At 
the same time, it is important to compare its performance with that observed in networks generated with both the ranking by rescaled indegree with random promotion, and the random algorithm. As pointed out above, for $\beta<1$, the nodes 
have a non-zero probability to choose their targets based on quality, which results in a non-zero indegree-quality even for networks generated with a random ranking. We focus on three values of $\beta$ ($\beta=0.7,0.8,0.9$) which correspond to populations of nodes that are mostly driven by ranking when selecting their targets; analogous results for smaller values of $\beta$ are shown in 
Supplementary Figs.~S16--S18 for $m=6$ and S8--S10 for $m=3$. 

We find (Fig.~\ref{kq_corr2}, top panels) that the indegree-quality correlation observed 
in $R(k)$-generated networks is not always larger than the indegree-quality correlation 
observed in random-generated networks: as the exploration cost $\alpha$ grows, the indegree-quality 
correlation in $R(k)$-generated networks dwindles; when $\alpha$ is larger than one, $R(k)$-generated
networks exhibit a smaller indegree-quality correlation than random-generated networks. While a large exploration cost is harmful for the overall indegree-quality correlation in both indegree-generated and $R(k)$-generated networks, for $\alpha\geq 1$, indegree's precision in promoting the top-quality nodes (Fig. \ref{kq_corr2}, bottom panels) tends to grow with the exploration cost. Remarkably, $R(k)$-generated networks exhibit the largest precision values for all the values of $\alpha$.

The behavior of the ($R(k)$+RP)-generated networks (i.e, the networks generated with the ranking by rescaled indegree with random promotion, $R(k)$+RP) is non-trivial. Indeed, for small values of $\alpha$, the indegree-quality correlation and indegree's precision of these networks are comparable with those of $R(k)$-generated networks, which indicates that randomly promoting a few nodes to the top of the ranking is not harmful to quality promotion. On the other hand, for $\alpha>1$, the indegree-quality correlation (or indegree's precision) of $R(k)$-generated networks becomes larger than that of ($R(k)$+RP)-generated networks. We conclude that for large values of $\beta$ ($\beta=0.7,0.8,0.9$), the random promotion mechanism has a marginal impact on quality promotion for sparse networks when $\alpha<1$, whereas it is harmful to quality promotion when the exploration cost is large, i.e., for $\alpha>1$. For smaller values of $\beta$ ($\beta=0.1,0.3,0.5$), the ($R(k)$+RP)-generated networks and the $R(k)$-generated networks exhibit comparable values of indegree-quality correlation and precision (Figs.~S8--S16).

\begin{figure*}[t]
\centering
\includegraphics[scale=0.4]{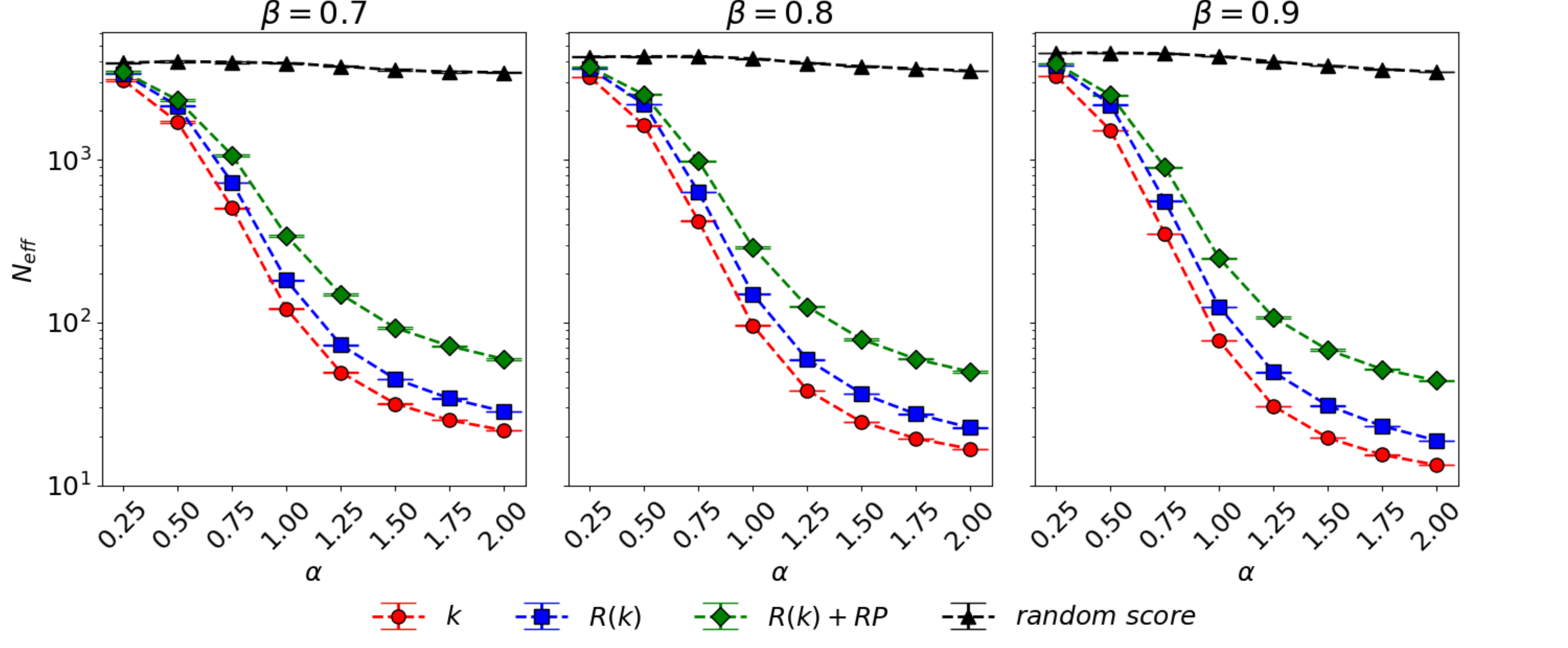}
\caption{Diversity as measured by $N_{eff}$(the larger, the more egalitarian the indegree distribution): comparison between indegree-generated (red circles), $R(k)$-generated (blue squares), $(R(k)+RP)$-generated (green rhombuses) and random-generated (black triangles) networks. The three columns correspond, from left to right, to $\beta=0.7,0.8,0.9$, respectively. The dots represent averages over $100$ realizations; the error bars represent the standard error of the mean.}
\label{h_index}
\end{figure*}

The qualitative difference between the top and the bottom panels of Fig.~\ref{kq_corr2} is explained
by the different indegree distributions of the networks generated with different values 
of $\alpha$. When $\alpha$ is large, the incoming links are concentrated on
few top items (as the effective number of nodes shows, see Fig.~\ref{h_index} below
and the related discussion) and, at the same time, the low-quality items remain unnoticed. The small number of incoming links received by low-quality items do not allow the metrics to discriminate their quality, which results in small indegree-quality correlation values.
By contrast, high-quality nodes receive a large number of incoming links, and it is possible for the metrics to rank them at the top, which results in relatively large precision values.

\begin{figure*}[t]
\centering
\includegraphics[scale=0.4]{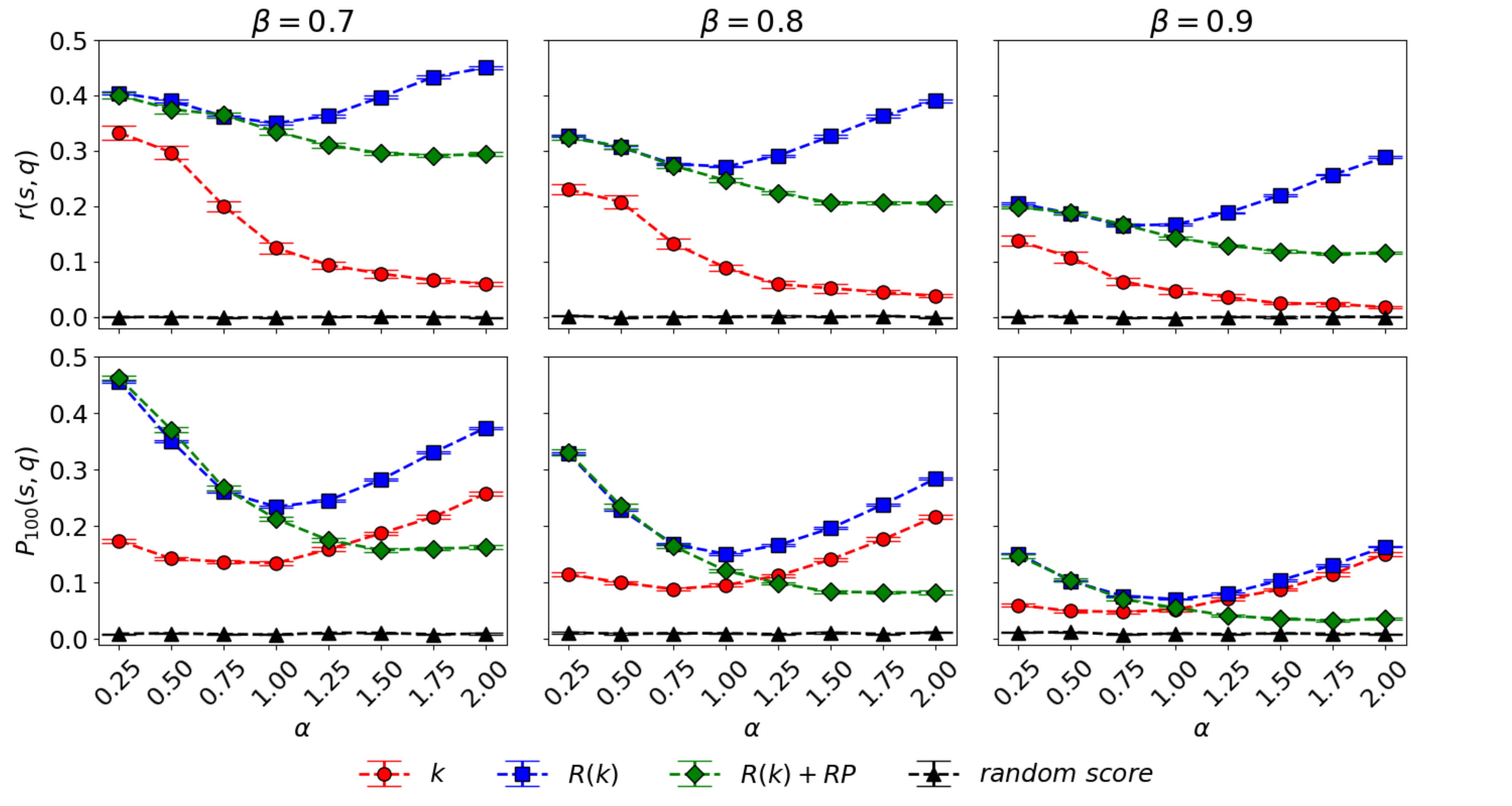}
\caption{Quality detection as measured by $r(s, q)$ (the Pearson's linear correlation between node score $s$ and node quality $q$ -- \emph{top} panels), and $P_{100}(s, q)$ (the precision of  node score $s$ in identifying the top-$100$ nodes by quality $q$ -- \emph{bottom} panels): comparison between indegree-generated ($s=k$, red circles), $R(k)$-generated ($s=R(k)$, blue squares), $(R(k)+RP)$-generated ($s=R(k)$, green rhombuses), and random-generated networks ($s$ is given by a random score, black triangles). The three columns correspond, from left to right, to $\beta=0.7,0.8,0.9$, respectively. The dots represent averages over $100$ realizations; the error bars represent the standard error of the mean. }
\label{sq_corr}
\end{figure*}

\subsection{Quality detection}
\label{sec:qd}
In indegree-generated and $R(k)$- generated networks, nodes' indegree and age-rescaled indegree, respectively, are the scores that are used for the nodes' ranking.
Our ability to detect quality in such networks is determined by the strength of the relation between node score $s$ and $q$ (as measured by both the Pearson's linear correlation $r^{s}(s,q)$ and the precision $P_{100}^{s}(s,q)$).
Remarkably, the correlation $r^R(R(k),q)$ is larger than the correlation $r^k(k,q)$ for all the parameter values (Fig.~\ref{sq_corr}, top panels, and Fig. S14). The precision of age-rescaled indegree in identifying the top-quality nodes is also larger than indegree's precision for all the parameter values (Fig.~\ref{sq_corr}, bottom panels). In qualitative agreement with the results obtained for quality promotion (Fig.~\ref{kq_corr}), for $m=6$ and $\beta=0.7,0.8,0.9$, the random promotion mechanism turns out to have a marginal impact for $\alpha<1$, whereas it is harmful for $\alpha>1$. For $m=3,6$ and $\beta=0.1,0.3,0.5$, the ($R(k)$+RP)-generated and $R(k)$-generated networks exhibit similar levels of score-quality correlation (see Figs. S9 and S17).

By being completely insensitive to node popularity, the random score always achieves zero precision, on average, in identifying the top-quality nodes. As expected, while the random score can still generate networks with non-zero indegree-quality correlation (Fig.~\ref{kq_corr2}), the rankings it produces have no practical utility.

\subsection{Diversity}
\label{sec:d}

For both indegree- and $R(k)$-generated networks, the effective number of nodes $N_{eff}$ depends on both $\alpha$ and $\beta$
(Figs. \ref{h_index}).
Unsurprisingly, the random score produces the most egalitarian networks (Fig.~\ref{h_index}), with $N_{eff}$ values above $3000=0.3\,N$.
In qualitative agreement with previous findings~\cite{salganik2006experimental},
the net effect of a larger sensitivity to the nodes' ranking position is, for all the studied metrics, a more unequal popularity distribution, which manifests itself in the decrease of $N_{eff}$ as $\alpha$ increases.
Importantly, the popularity distribution is evener for $R(k)$-generated networks
than for indegree-generated networks (Figs.~\ref{h_index} and S15), and the ($R(k)$+RP) algorithm further enhances popularity diversity.
In summary, the age normalization procedure not only improves the indegree-quality and the score-quality correlation,
but it also decreases the popularity inequality in the system; as expected, the random promotion mechanism further decreases the popularity inequality.
At the same time, both the $R(k)$-generated and the ($R(k)$+RP)-generated networks exhibit $N_{eff}$ values significantly smaller than the $N_{eff}$
achieved by the random ranking. It remains open to design ranking algorithms that lead to more egalitarian networks than those generated with rescaled indegree and ($R(k)$+RP),
yet maintaining a similar
level of quality promotion and quality detection.

\subsection{Including node removal in the model}
\label{sec:r}

\begin{figure*}[t]
\centering
\includegraphics[scale=0.4]{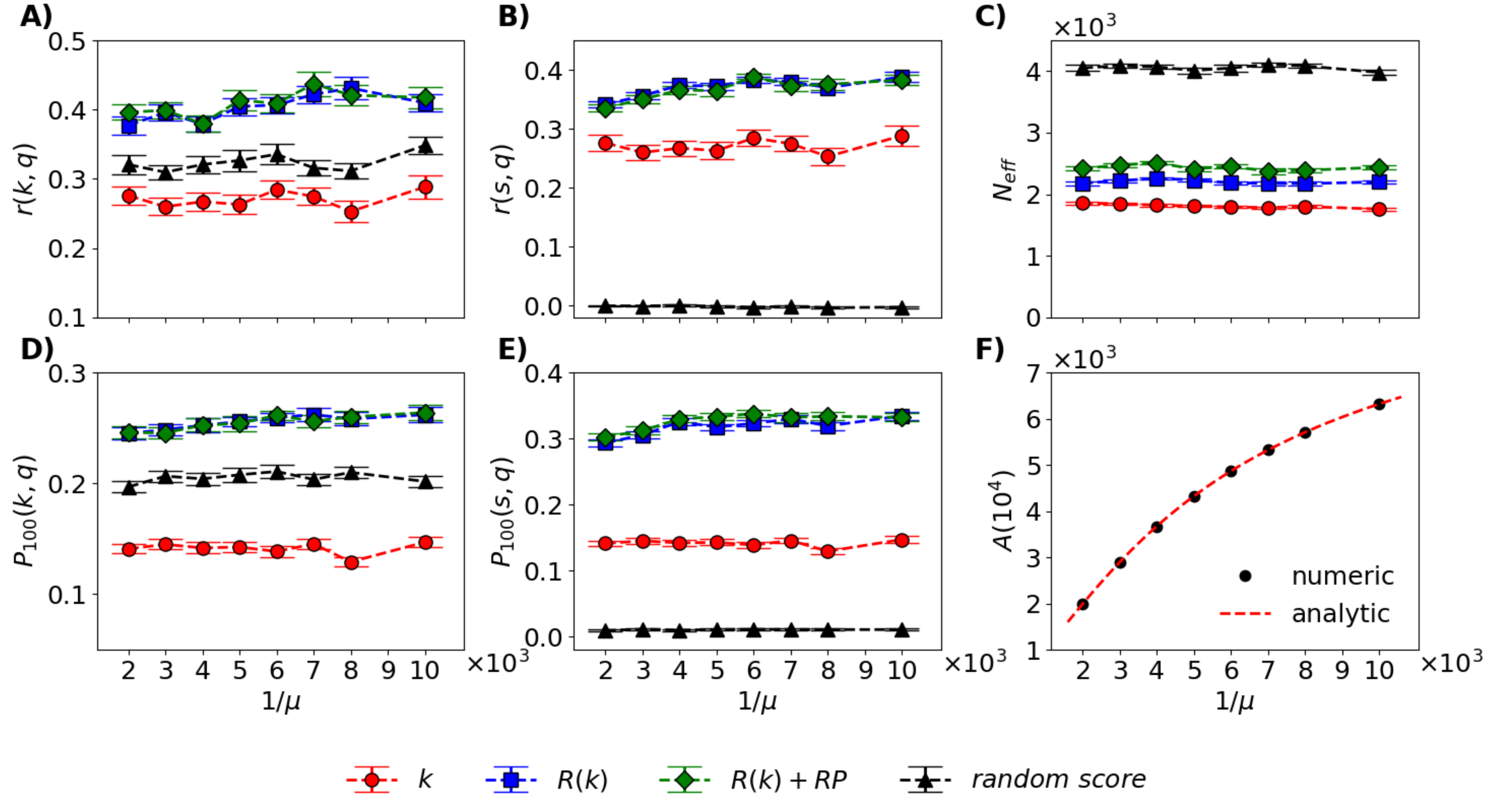}
\caption{The impact of node removal on quality promotion, quality detection, popularity diversity. 
We grow networks generated with the model described in Section~\ref{sec:r} with $\alpha=0.5$, $\beta= 0.7$, $T=10^{-4}$. We show five ranking evaluation metrics as a function of the inverse of the removal probability, $A^{*}=\mu^{-1}$: (A) the Pearson's correlation between node indegree and node quality; (B) the Pearson's correlation between node score and node quality (see Fig.~\ref{sq_corr}'s caption for the definition of node score); (C) the effective number of nodes, $N_{eff}$; (D) the indegree's precision in identifying the top-$100$ nodes by quality; (E) the score's precision in identifying the top-$100$ nodes by quality. As in the previous figures, different lines correspond to the networks generated with different algorithms. The symbols represent averages over $50$ realizations; the error bars represent the standard error of the mean. Panel F shows the number of active nodes at time $t=10^{4}$, i.e., at the time when we halt the simulations, as a function of $\mu^{-1}$; the values computed analytically through Eq.~\eqref{expected} well match the values observed in the simulations.}
\label{fig:removal}
\end{figure*}

In real social and information systems, not only new nodes can enter the system, but also existing nodes can disappear or lose relevance. The members of an online community, for example, may lose interest in the platform and deactivate their account~\cite{dror2012churn}, which may eventually lead to the ``death" of the platform~\cite{garcia2013social}. In the WWW, many webpages are deleted every day, which makes it essential to incorporate node deletion into growing network models~\cite{kong2008experience}. To take into account this situation, we introduce a variant of the model introduced in Section~\ref{sec:model} where the nodes are in one of two possible states: ``active" and ``removed". At every time step $t$, each active node becomes removed with probability $\mu$, and one new active node enters the system and chooses its targets among the existing active nodes according to the rules described in Section~\ref{sec:model}. In the continuum approximation, the number of active nodes, $A(t)$, follows the equation $\dot{A}(t)=1-\mu\,A(t)$ whose solution (with the initial condition $A(1)=1$) is 
  \begin{equation}
  A(t)=\mu^{-1}+(1-\mu^{-1})\,\exp{(-\mu\,(t-1))}.
  \label{expected}
  \end{equation}
  After an initial linear growth ($A(t)\approx t$ for $t\ll \mu^{-1}$), the number of active nodes eventually equilibrates at $A^*=1/\mu$. The validity of Eq.~\eqref{expected} is confirmed by our numerical simulations (see Fig.~\ref{fig:removal}F). Note that
  when $\mu=0$ (i.e., in the model without node removal), $A(t)=t$ and $N=t^*$, where $t^*$ denotes the total number of performed simulation steps. In the following, we set $t^*=10^{4}$; due to the node removal mechanism, the number $A(10^4)$ of active nodes at the end of the simulation is substantially smaller than $10^4$, as expected based on Eq.~\eqref{expected} (see Fig.~\ref{fig:removal}F).

We investigated whether this node removal mechanism alters substantially the results described above. More specifically, we performed numerical simulations with values of $\mu$ ranging from $10^{-4}$ to $5\cdot 10^{-4}$, which corresponds to expected values of the asymptotic number of active nodes, $A^*=1/\mu$, ranging from $2,000$ to $10,000$. We find that through the whole range of $\mu$ values, the results are qualitatively similar to those obtained for the networks without node removal (see Figs.~\ref{fig:removal}A-E for the results for $\alpha=0.5,\beta=0.7$, and Figs.~S25 for $\alpha=1,\beta=0.7$). All the considered observables display a remarkable stability with respect to $\mu$, which indicates that node removal has a marginal impact on the algorithms' quality promotion, quality detection, and popularity diversity. It remains open to determine the largest removal probability $\mu$ tolerated by the system before the results change qualitatively.

\subsection{A case study: growing a network of scientific papers based on different metrics}
\label{sec:aps}

So far, we have considered model-generated networks.
How are our model and our results on ranking algorithms relevant to real systems?
If our model provides a plausible description of the growth of a given system of interest,
and we have a reliable way to infer the ($\alpha$, $\beta$) parameters from the available data, we would be able to quantify the potential impact of various ranking algorithms on the future properties of the system.
Stimulated by this observation, in this Section, we analyze real information networks to fit our model's parameter to the empirical data, and use the resulting parameters to grow again the network based on different ranking algorithms.  

The information networks analyzed here are subsets of the American Physical Society (APS) citation network of scientific papers\footnote{The dataset has been already analyzed in~\cite{medo2011temporal,mariani2016identification,mariani2018early}, and it can be downloaded here: \url{https://journals.aps.org/datasets}.}. The citation dataset provided by the APS contains all $539974$ papers published by the APS from 1893 to 2013 together with their $5992897$ references to other papers published by APS journals and their publication date. For our analysis, we extract two subsets of papers that include the papers with the PACS number\footnote{The PACS (Physics and Astronomy Classification Scheme) codes refer to a hierarchical classification of research areas in physics, astronomy, and related sciences. We refer to~\url{https://journals.aps.org/PACS} for further information.} 89.75.Hc (``Networks and genealogical trees"; the subset includes $N=1615$ papers and $L=8222$ citations) and 03.67.Lx (``Quantum computation architectures and implementations"; the subset includes $N=3876$ papers and $L=24213$ citations), respectively.

To quantify the potential impact of different ranking algorithms, the first step is to infer the optimal pair $(\hat{\alpha},\hat{\beta})$ of model parameters together with the optimal algorithm that together lead to the best agreement between the $\mathcal{A}$-generated model networks and the observed real network. The real data determine the final number of nodes in the network, their order of appearance, and their outdegree values; in an $\mathcal{A}$-generated model network, the nodes choose their references based on the rules of the model described in Section~\ref{sec:model}, where the ranking position of the nodes is determined by the chosen algorithm $\mathcal{A}$. 

Quality is not accessible in real data as opposed to synthetic networks where it is a well-defined node-level variable. To overcome this limitation, as rescaled indegree is the best-performing metric in quantifying node quality in synthetic networks (Fig.~\ref{sq_corr}), we use the papers' final rescaled indegree score $R(k)$ as a proxy for their quality if $R(k)\geq 0$, whereas we set $q=0$ for papers such that $R(k)<0$.

The agreement between the $\mathcal{A}$-generated networks and the original network is quantified by the model's mean error per node, $e^{(\mathcal{A},\alpha,\beta)}$. For each $\mathcal{A}$-generated network\footnote{An $\mathcal{A}$-generated network is a realization of the stochastic process that generates $\mathcal{A}$-generated networks.} $\mathcal{G}(\mathcal{A},\alpha,\beta)$, we first define the network's mean error per node, $e(\mathcal{G}(\mathcal{A},\alpha,\beta))$, as
\begin{equation}
e(\mathcal{G}(\mathcal{A},\alpha,\beta))=\frac{1}{N}\sum_{i=1}^{N}|k_i(\mathcal{G}(\mathcal{A},\alpha,\beta))-k_i^*|,
\end{equation}
where $k_i(\mathcal{G})$ denotes paper $i$'s indegree in network $\mathcal{G}$, whereas $k_i^*$ denotes the observed indegree of paper $i$.
The model's mean error per node, $e^{(\mathcal{A},\alpha,\beta)}$ is defined as the average of $e(\mathcal{G})$ over a sample of 100 $\mathcal{A}$-generated networks.
Essentially, the model's mean error per node quantifies the average deviation between node indegree in real and model networks. The lower $e^{(\mathcal{A},\alpha,\beta)}$, the better the agreement between the $\mathcal{A}$-generated model networks and the real network.

Finally, we compare the properties of the networks generated by different ranking algorithms for $(\alpha,\beta)=(\hat{\alpha},\hat{\beta})$. 
The rationale behind such a comparison is that the parameter values $(\hat{\alpha},\hat{\beta})$ that yield the best agreement represent our best estimations of the nodes’ exploration cost $\alpha$ and sensitivity to popularity $\beta$; based on this assumption, we estimate the long-term implications
of various algorithms by simply changing the algorithm that is used to grow the
network, whilst keeping $\alpha=\hat{\alpha}$ and $\beta=\hat{\beta}$ fixed.

\begin{figure*}[t]
\centering
\includegraphics[scale=0.55]{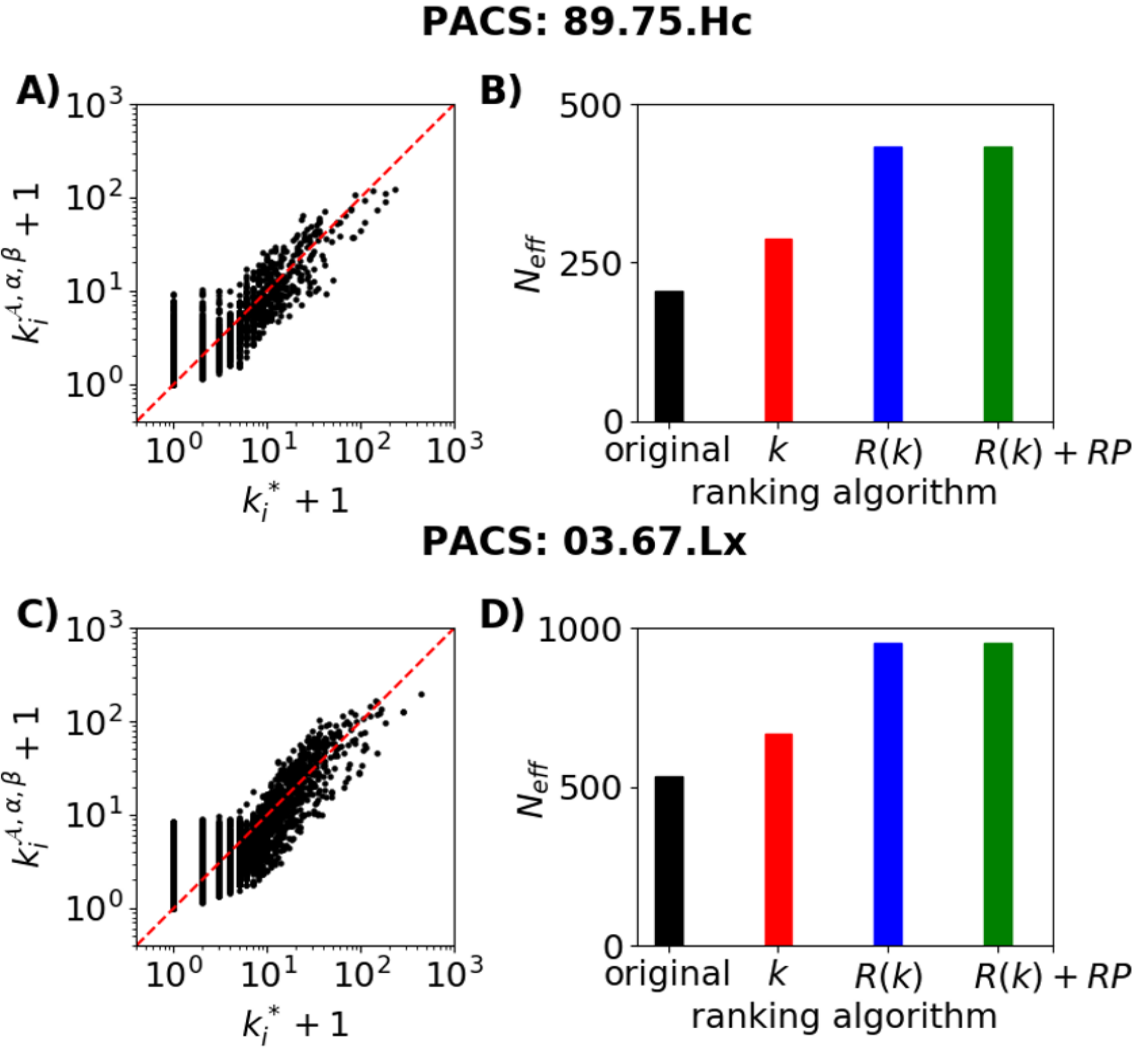}
	\caption{A comparison of two real networks (subsets of the APS citation network) with the respective synthetic networks generated with the model described in Section~\ref{sec:model}, using different ranking algorithms. Panels (A,C) illustrate the scatter plots between the nodes' original indegree $k^*$ and the nodes' average indegree $\overline{k}$ in indegree-generated networks ($\alpha=\hat{\alpha}, \beta=\hat{\beta}$) generated as described in the main text, for the APS network's subset that corresponds to the PACS code 89.75.Hc and 03.67.Lx, respectively. Panels (B,D) show the averages of $N_{eff}$ over $100$ realizations of the model networks generated by different ranking algorithms. The ($R(k)$+RP)-generated ($\eta=0.5, T=10$) and $R(k)$-generated networks exhibit significantly larger values of $N_{eff}$ than the original network and the indegree-generated network. This indicates that in a real information system, suppressing the age-bias of popularity-based metrics can substantially improve popularity diversity.}
    \label{fig:aps}
\end{figure*}

For the subset that corresponds to the PACS code 89.75.Hc, we find that indegree-generated networks with $\hat{\alpha}=0.5$ and $\hat{\beta}= 0.5$ exhibit the best agreement with the real network (mean error per node $e^{(k,0.5,0.5)}=3.058$, see Fig. (see Fig.~\ref{fig:aps}A)).
Therefore, we fix $\alpha=0.5$ and $\beta=0.5$, and we compare the networks generated by three different algorithms: indegree, age-rescaled indegree, and the ranking by rescaled indegree with random promotion.
In qualitative agreement with our previous results, we find that $R(k)$-generated networks and ($R(k)$+RP)-generated networks exhibit substantially larger popularity diversity than $k$-generated networks and the original network (see Fig.~\ref{fig:aps}B). Qualitatively similar results are obtained for the subset that corresponds to the PACS code 03.67.Lx (see Figs.~\ref{fig:aps}C-D). These results confirm that in a real information system, suppressing the age bias of popularity metrics can increase popularity diversity and reduce the mean age of the most popular nodes in the system. Beyond numerical simulations, we envision that future studies might further validate this conclusion by means of field experiments where subjects in different groups can select various information items based on the items' ranking position by different ranking algorithms.

\section{Discussion}
\label{sec:discussion}

To summarize, we find that age-rescaled indegree allow us not only to fairly compare old and recent nodes \cite{newman2009first,mariani2016identification}, but
also to produce networks where the nodes' final popularity is 
better correlated with their quality than for the networks that adopted indegree, and the popularity distribution is more even.
Examples of widely-used cumulative popularity metrics include the number of views or downloads for online content, the number of received citations for scientific papers, among others.
Our results indicate that despite the widespread use of cumulative popularity metrics, age-rescaled metrics may better help both users to find high-quality content, and high-quality content to experience larger success than low-quality content.
Besides, the random promotion mechanism introduced here consistently improves popularity diversity, whereas its impact on quality promotion and detection ranges from marginal to substantial depending on the exploration cost parameter $\alpha$.

The main message of our work is that 
network-based growth models can help us not only to understand the impact of network growth mechanisms on the rankings by a given algorithm~\cite{mariani2015ranking,medo2016model,liao2017ranking}, but also to estimate the impact of the adoption of different ranking algorithms
by a given system. In other words, we can investigate not only how the past evolution of the system influenced the current rankings, but also how the adopted rankings
may influence the future evolution of the system. As a result, our ranking-driven and quality-driven growing network model can be interpreted as a generative model for benchmark graphs to evaluate the performance of ranking algorithms in terms of quality promotion, quality detection, and popularity diversity.

We stress that the ranking algorithms considered here are simple in the sense of being based on a single criterion that is furthermore readily quantified; consider the ranking of nodes by their indegree, for example. The newly proposed ranking algorithm with random promotion is similar as it only requires choosing the original metric and the number of promoted nodes in the top of the ranking. In a general case where multiple criteria are to be considered, especially when they are contradicting such as node popularity and novelty, the existing vast body of literature on multiple-criteria decision analysis~\cite{triantaphyllou2000multi,greco2016multiple} with techniques such as the analytic hierarchy process~\cite{saaty2013analytic} and outranking~\cite{roy1990outranking} becomes relevant. This goes beyond directly combining node scores by various metrics~\cite{okamoto2008ranking,zhou2010solving} that is often used in the field of network analysis, yet it can become relevant when designing a system for real users with their heterogeneous, and often contradictory, needs and preferences.

Our work sheds light on the long-studied relation between quality/talent and success: Do the high-quality nodes experience larger success than the low-quality nodes? Why nodes of similar worthiness experience widely different success? In real systems, addressing these questions is challenging as defining ``node quality" in an unbiased and objective way is often not possible.
Our model-based approach bypasses this obstacle by defining node quality as an intrinsic node property,
and by building multiple independent realizations of an artificial system where the nodes choose their connections based on both the other nodes' ranking and their quality.
At the same time, while the model studied here is arguably one of the simplest models which feature all the elements of interest
in our analysis (network growth, and the joint influence of ranking and quality on network growth), it can only provide a stylized description of the growth of real networks.

Finally, the application of our growing network model to the American Physical Society citation network of scientific papers constitutes an attempt to estimate the potential consequences of various ranking algorithms on a given information system. We envision that more sophisticated models together with suitable field experiments will improve the reliability of model-based predictions of
the effects of ranking algorithms, providing us with a robust basis for more informed choices of ranking algorithms for real-world applications.
To draw a parallel, in a similar way as high-resolution models of epidemic spreading have led to accurate predictions of the properties of disease outbreaks~\cite{tizzoni2012real}, detailed models of network evolution may lead to the accurate quantification of the consequences of the adoption of a given metric in a given system.

\section*{Acknowledgments}
We thank Yi-Cheng Zhang for many enlightening discussions on the topic. This work has been supported by the National Natural Science Foundation of China (Grants Nos. 61673150, 11622538), the Science Strength Promotion Program of the UESTC, and the 
Zhejiang Provincial Natural Science Foundation of China (Grant no. LR16A050001). MSM acknowledges the University of Z\"urich for support through the URPP Social Networks. 

\section*{Author contributions statement}

M.S.M. and M.M. conceived the idea, M.S.M. and L.L. designed research,  S.Z. performed the numerical simulations, S.Z. and M.S.M. performed the analytic computations, all authors analyzed and discussed the results. S.Z. and M.S.M. wrote the manuscript. All authors reviewed the manuscript. 

\bibliographystyle{plainnat}

\appendix

\section{Details on the numerical simulations}
\label{sec:details}

We focus on networks composed of $10^4$ nodes, and study the following model parameters: $\alpha$ from $0.25$ to $2.0$ with step $0.25$ and $\beta$ from $0$ to $1$ with step $0.1$. Node quality values are drawn from the Pareto distribution $P(q)\sim q^{-3}$ where $q\in[1,\infty)$. The network is initialized with a network of $m$ nodes, each of them with one outgoing and incoming link. At each time step $t>m+1$, a new node $t$ is added to the system and $m$ nodes (only results for $m=6$ are shown in main text, whereas results for both $m=3$ and $m=6$ are shown in the Supplementary Material) are chosen as targets to establish $m$ new links. With probability $\beta$, the probability that a given node is chosen is given by Eq.~\eqref{ranking}, with probability $1-\beta$, it is given by Eq.~\eqref{quality}. To save computational time, for times $t\le10^2$, we re-compute and update the rankings at each time step, whereas for times $t\ge10^2$, the newly introduced nodes are placed at the bottom of the node ranking, and we re-compute and update the rankings every $10$ time steps. To make our results insensitive to random fluctuations, for each parameter pair $(\alpha, \beta)$, all the results shown here represent averages over a sufficiently large number of realizations of the network growth process.

\section{The relation between popularity and quality in indegree-generated networks}
\label{sec:relation}

We start by considering networks where the nodes cannot perceive the other nodes ranking, and are completely driven by quality ($\beta=1$).
In this scenario, the average indegree of node $i$ at time $N$ is given by
\begin{eqnarray}
\overline{k_{i}}(N)=\sum_{t=i+1}^{N}m\,\frac{q_i}{\sum_{j=1}^{t-1}q_j}.
\end{eqnarray}
In the thermodynamic limit $N\gg i$, by using a similar mean-field approximation as in~\cite{fortunato2006scale},
we obtain
\begin{eqnarray}
\overline{k_{i}}(N)\simeq m\,\frac{q_i}{\overline{q}}\,\log{\Biggl(\frac{N-1}{i-1} \Biggr)}.
\label{onlyq}
\end{eqnarray}
There is a good agreement between Eq.~\eqref{onlyq} and the results of numerical simulations (see Supplementary Fig.~S26).
Such linear relation between indegree and quality does not hold for $\beta>0$, where the analytic calculation is made difficult by the fact that the ranking position of a given node at a given time is influenced by both its quality and the previous dynamics of the system. Nevertheless, we find that the relation $\overline{k_i}(N)=C\,q_i^{\delta}$ fits reasonably well the simulation results, and the dependence of the fitted exponent $\delta$ on node age is relatively weak (see Fig.~S27 and Table~S1 for details).


\pagebreak
\begin{center}
\textbf{\large Supplemental Materials: \\
The long-term impact of ranking algorithms in growing networks
}
\end{center}

\setcounter{table}{0}
\renewcommand{\thetable}{S\arabic{table}}
\setcounter{figure}{0}
\renewcommand{\thefigure}{S\arabic{figure}}

\begin{figure*}[h]
\centering
\includegraphics[scale=0.3]{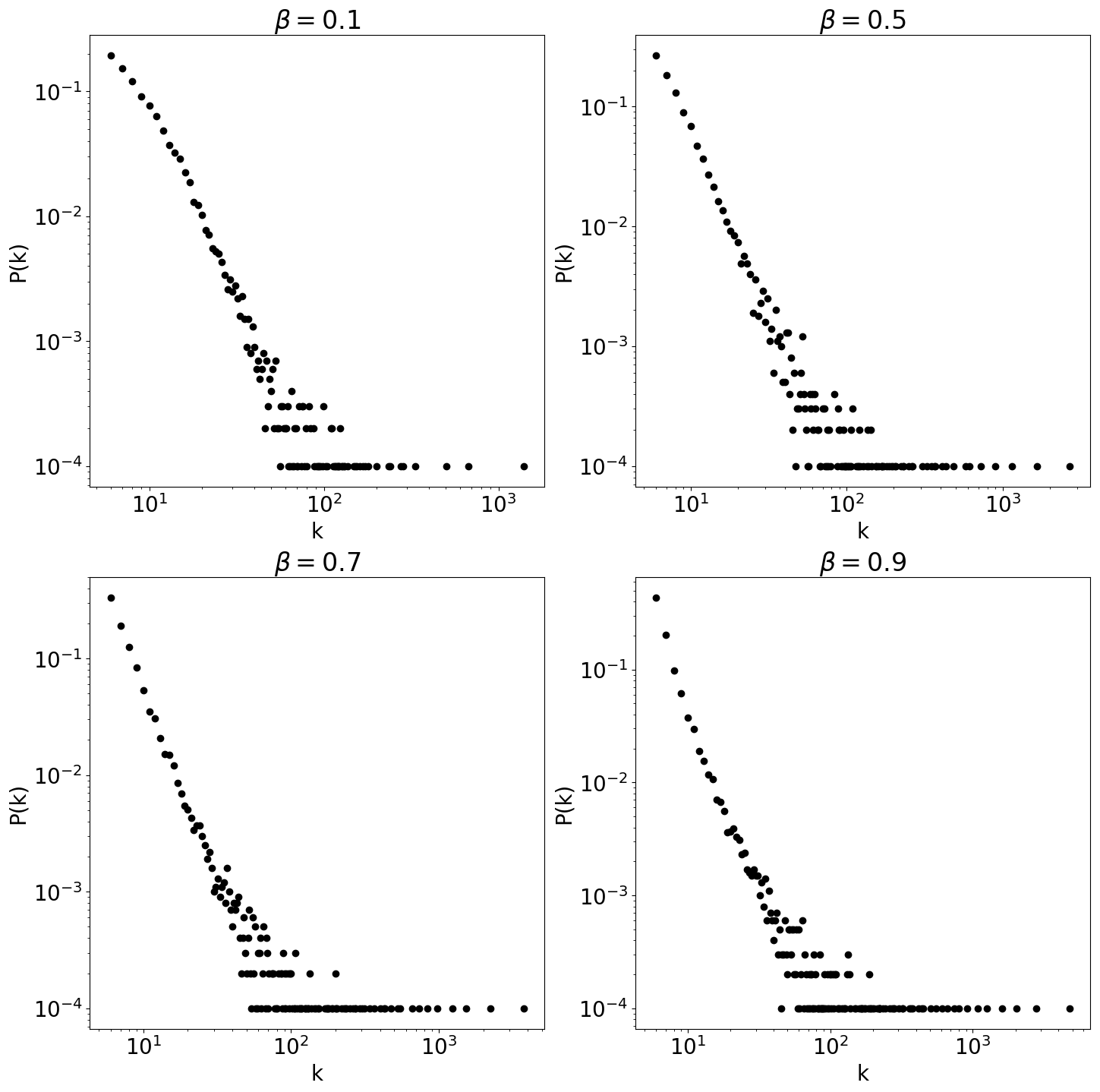}
	\caption{Results for $m=6$. The degree distribution of indegree-generated networks, the four panels correspond to  for $\beta=0.1, 0.5, 0.7, 0.9$ with fixed $\alpha=1.0$ respectively.}
\end{figure*}


\begin{figure*}[h]
\centering
\includegraphics[scale=0.3]{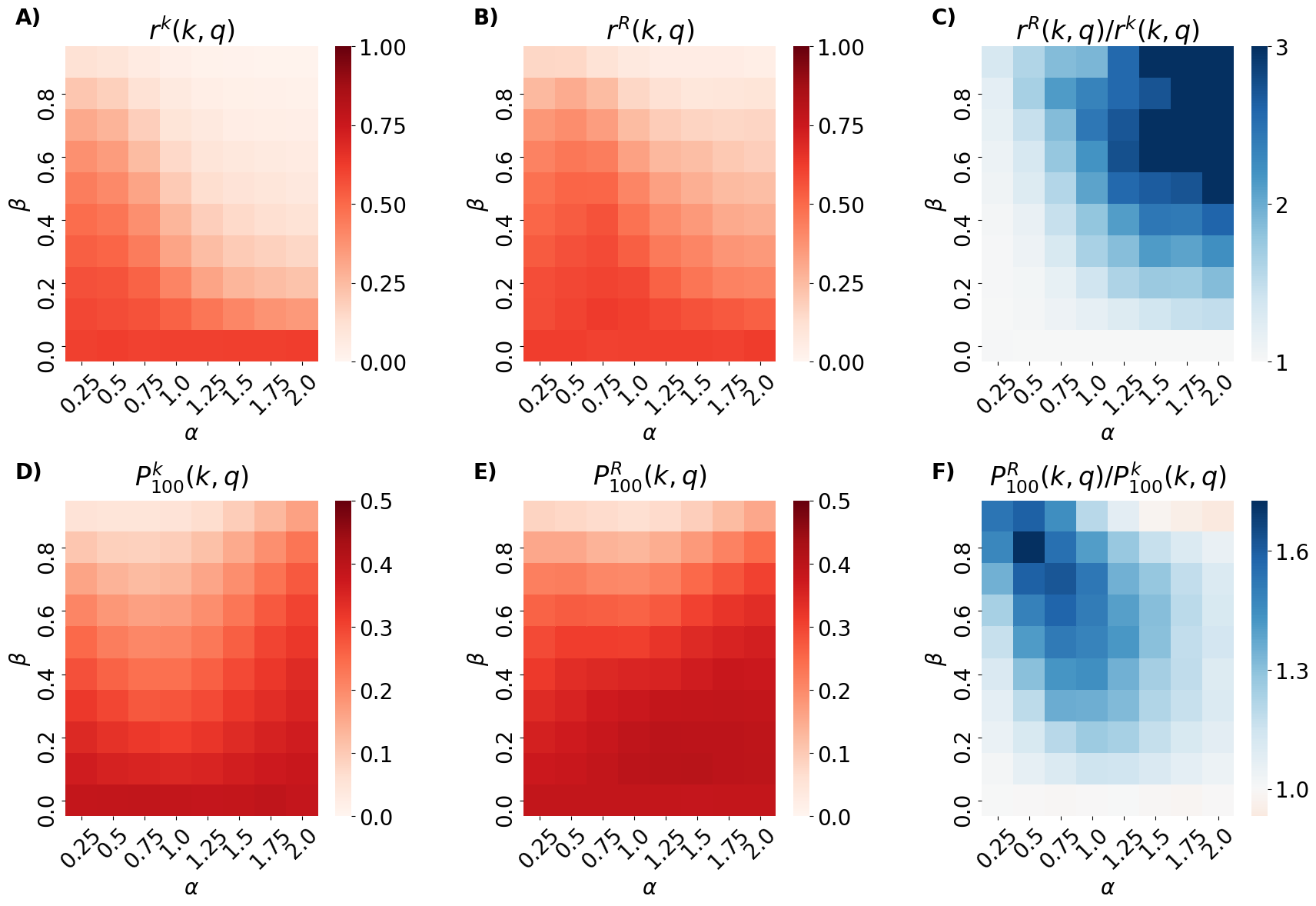}   
	\caption{Results for $m=3$. Quality promotion as measured by $r(k,q)$ (the Pearson's linear correlation between node indegree $k$ and node quality $q$), and $P_{100}(k, q)$ (the precision of node indegree $k$ in identifying the top-100 nodes by quality $q$ -- \emph{bottom} panels): comparison between indegree-generated and $R(k)$-generated 
networks. (A-B): $r(k, q)$ for indegree-generated ($r^k(k, q)$, panel A) and $R(k)$-generated ($r^R(k, q)$, panel B) networks,
as a function of the model's parameter $\alpha$ (exploration cost) and $\beta$ (reliance on ranking). (C): Ratio $r^{R}(k,q)/r^{k}(k,q)$ as a function of the model's parameters. (D-E): $P_{100}(k, q)$ for indegree-generated ($P^k_{100}(k, q)$, panel D) and $R(k)$-generated ($P^R_{100}(k, q)$, panel E) networks, as a function of the model's parameters. (F): Ratio $P^R_{100}(k, q)/P^k_{100}(k, q)$ as a function of the model's parameters. Results are averaged over $500$ realizations.}
\end{figure*}


\begin{figure*}[h]
\centering
\includegraphics[scale=0.3]{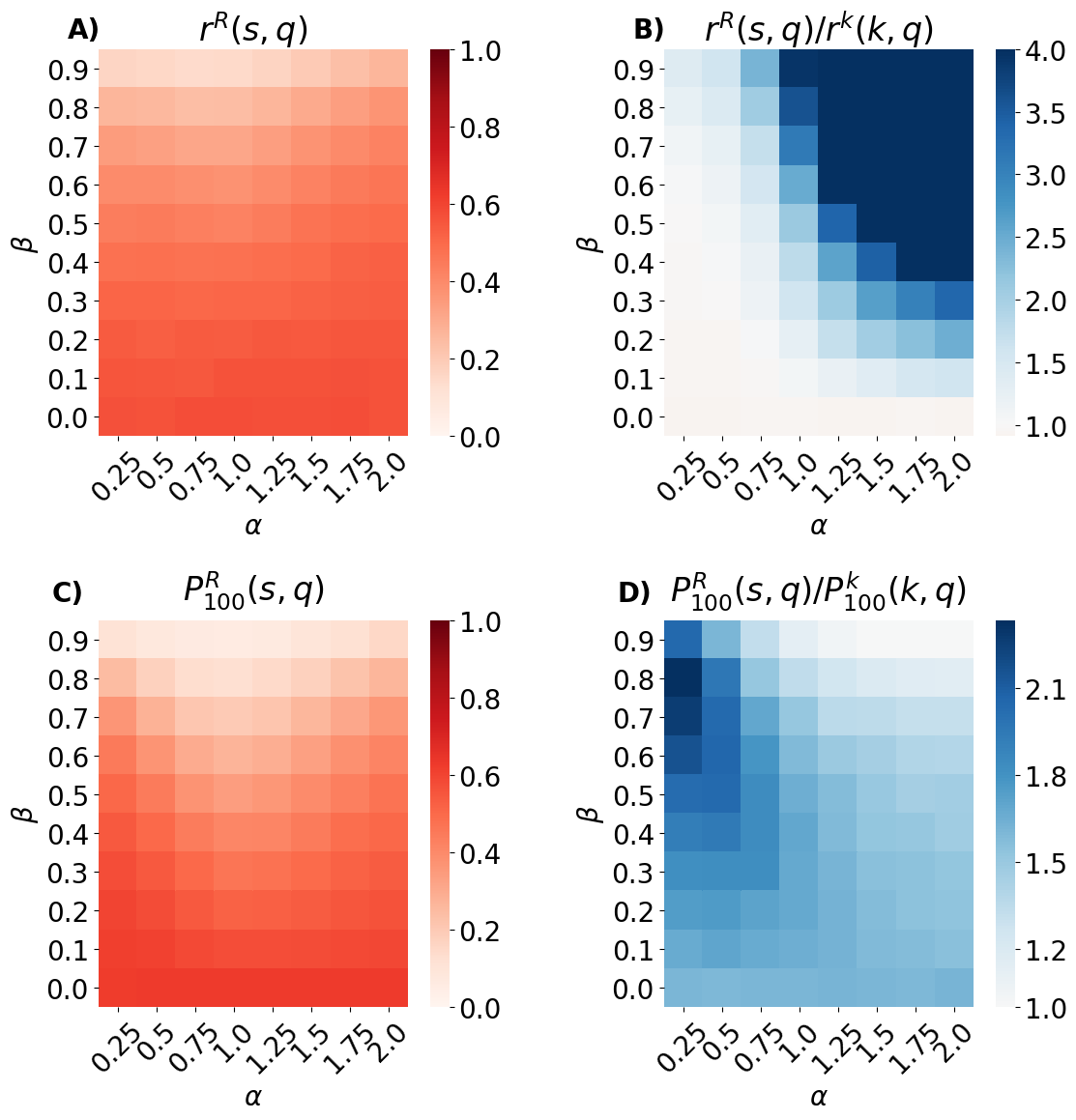}
	\caption{Results for $m=3$. Quality detection as measured by $r(s, q)$ (the Pearson's linear correlation between 
node score $s$ and node quality $q$ -- \emph{top} panels), and $P_{100}(s, q)$ (the precision of node score $s$ in identifying the top-100 nodes by quality $q$ -- \emph{bottom} panels): comparison between indegree-generated and $R(k)$-generated networks. (A) $r(s, q)$ for $R(k)$-generated networks, as a function of the model's parameter $\alpha$ (exploration cost) and $\beta$ (popularity bias). (B) Ratio $r^{R}(s,q)/r^{k}(k,q)$ as a function of the model's parameters. (C) $P_{100}(s,q)$ for $R(k)$-generated networks, as a function of the model's parameters. (D) Ratio $P_{100}^{R}(s,q)/P_{100}^{k}(k,q)$ as a function of the model's parameters. Results are averaged over $500$ realizations.}
\end{figure*}

\begin{figure*}[h]
\centering
\includegraphics[scale=0.3]{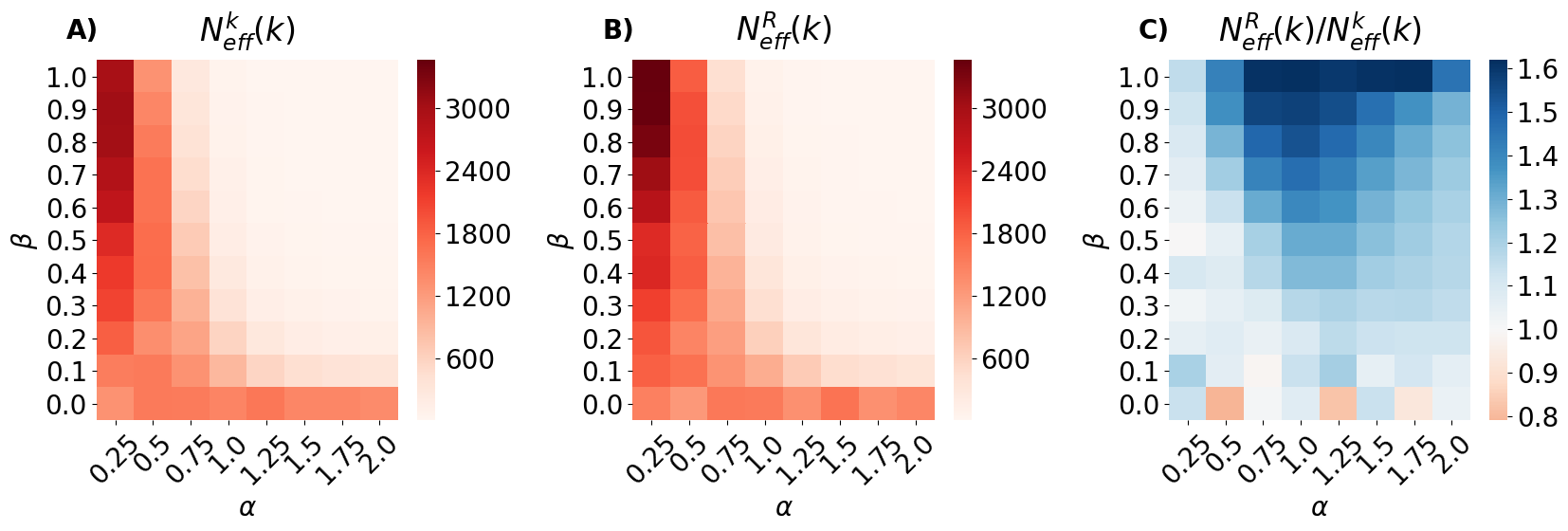}
	\caption{Results for $m=3$. Diversity as measured by $N_{eff}$(the larger, the more egalitarian the indegree distribution): comparison between popularity-generated and $R(k)$-generated networks. (A-B) The effective number of nodes $N_{eff}(k)$ for indegree-generated ($N_{eff}^k(k)$, panel A) and $R(k)$-generated ($N_{eff}^R(k)$, panel B) networks, as a function of the model's parameters. (C) Ratio $N_{eff}^R(k)/N_{eff}^k(k)$ as a function of the model's parameters. Results are averaged over $500$ realizations.}
\end{figure*}


\begin{figure*}[h]
\centering
\includegraphics[scale=0.4]{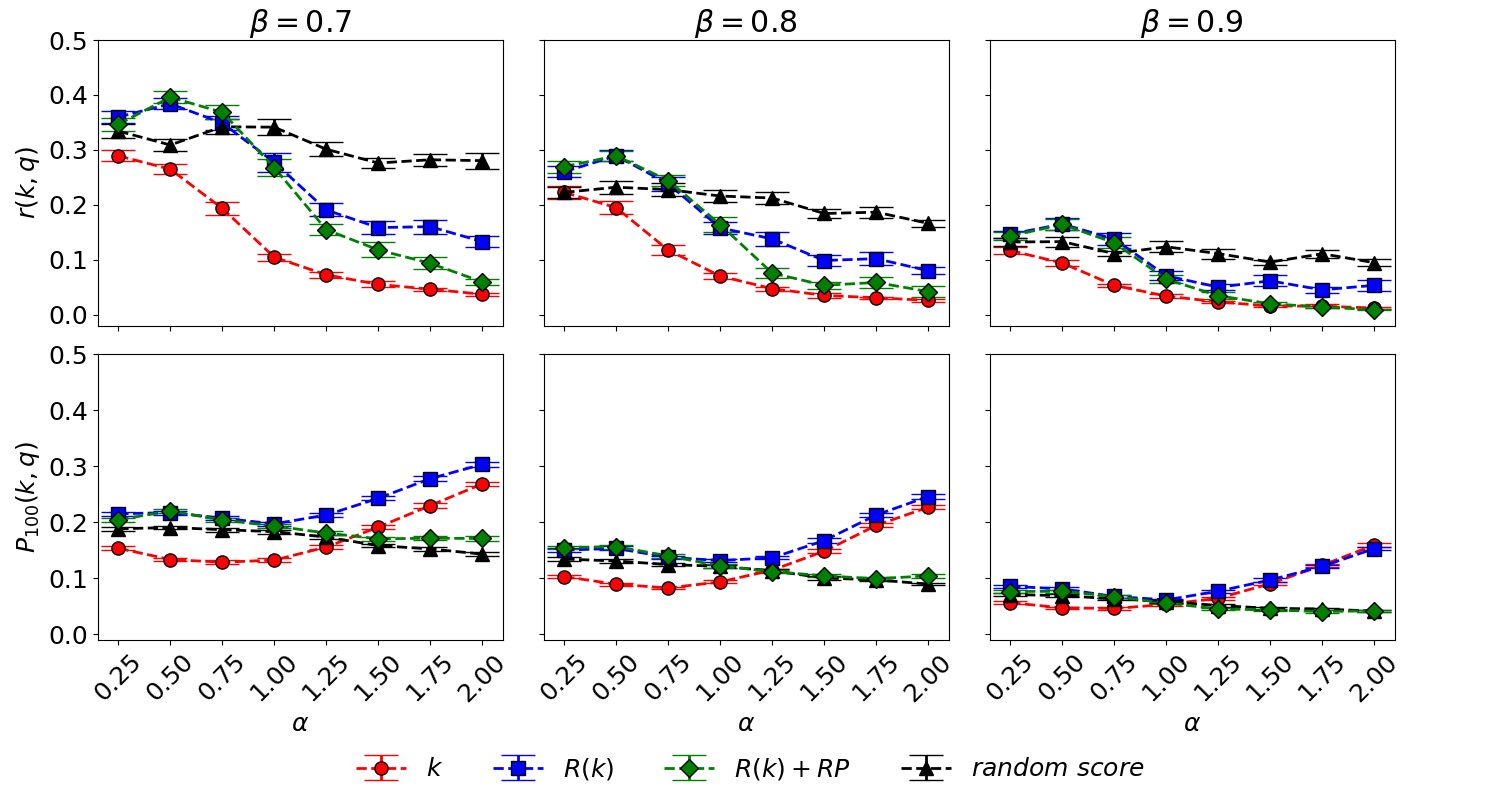}
	\caption{Results for $m=3$. Quality promotion as measured by $r(k, q)$ (the Pearson's linear correlation between node indegree $k$ and node quality $q$ -- \emph{top} panels), and $P_{100}(k, q)$ (the precision of node indegree $k$ in identifying the top-$100$ nodes by quality $q$ -- \emph{bottom} panels): comparison between indegree-generated (circles), $R(k)$-generated (squares), $R(k)+RP$-generated (green rhombuses), and random-generated networks (triangles). The three columns correspond, from left to right, to $\beta=0.7,0.8,0.9$, respectively. Results are averaged over $100$ realizations; the error bars represent the standard error of the mean.}
\end{figure*}

\begin{figure*}[h]
\centering
\includegraphics[scale=0.4]{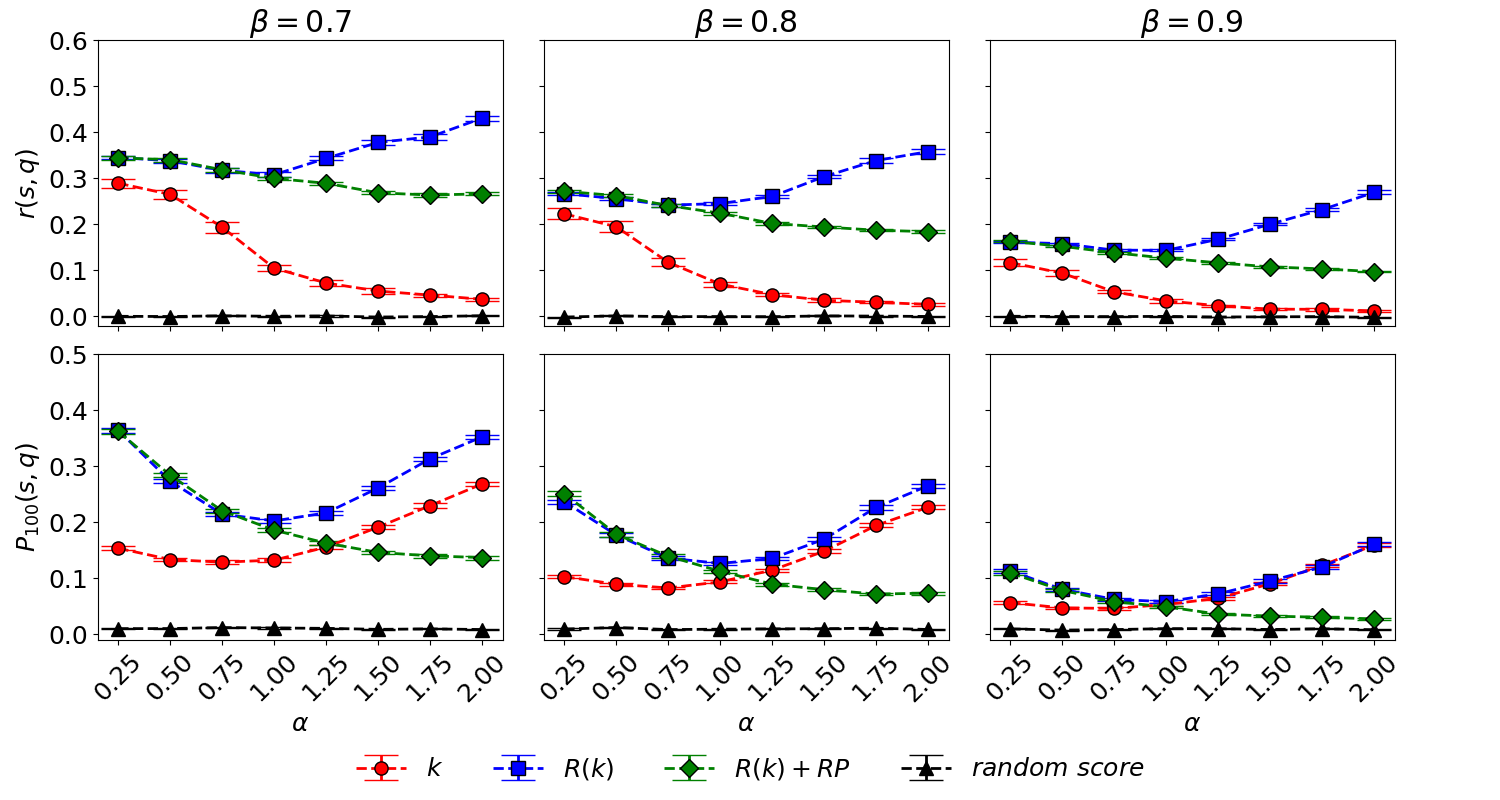}
	\caption{Results for $m=3$. Quality detection as measured by $r(s, q)$ (the Pearson's linear correlation between node score $s$ and node quality $q$ -- \emph{top} panels), and $P_{100}(s, q)$ (the precision of  node score $s$ in identifying the top-$100$ nodes by quality $q$ -- \emph{bottom} panels): comparison between indegree-generated ($s=k$, circles), $R(k)$-generated ($s=R(k)$, squares), $R(k)+RP$-generated (green rhombuses), and random-generated networks ($s=\rho$, triangles). The three columns correspond, from left to right, to $\beta=0.7,0.8,0.9$, respectively. The dots represent averages over $100$ realizations; the error bars represent the standard error of the mean. }
\end{figure*}


\begin{figure*}[h]
\centering
\includegraphics[scale=0.4]{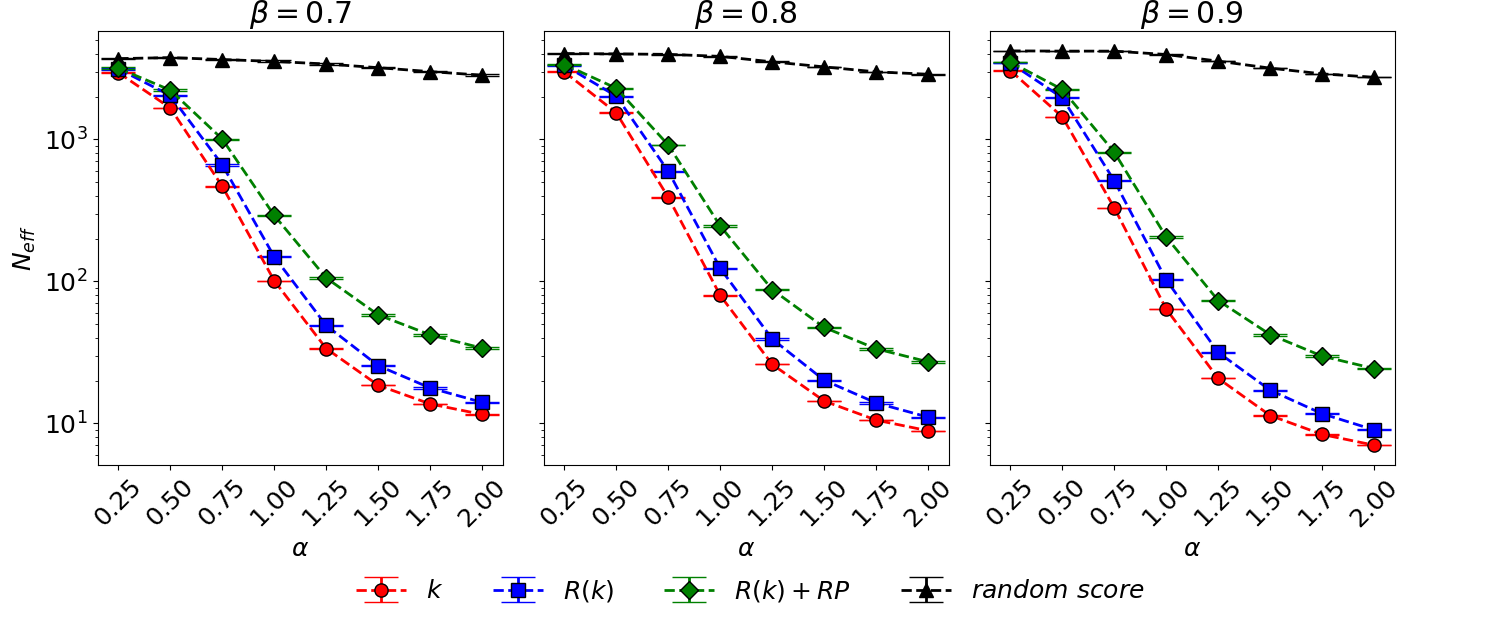}
	\caption{Results for $m=3$. Diversity as measured by $N_{eff}$(the larger, the more egalitarian the indegree distribution): comparison between indegree-generated (circles), $R(k)$-generated (squares), $R(k)+RP$-generated (green rhombuses), and random-generated (triangles) networks. The three columns correspond, from left to right, to $\beta=0.7,0.8,0.9$, respectively. The dots represent averages over $100$ realizations; the error bars represent the standard error of the mean.}
\end{figure*}

\begin{figure*}[h]
\centering
\includegraphics[scale=0.4]{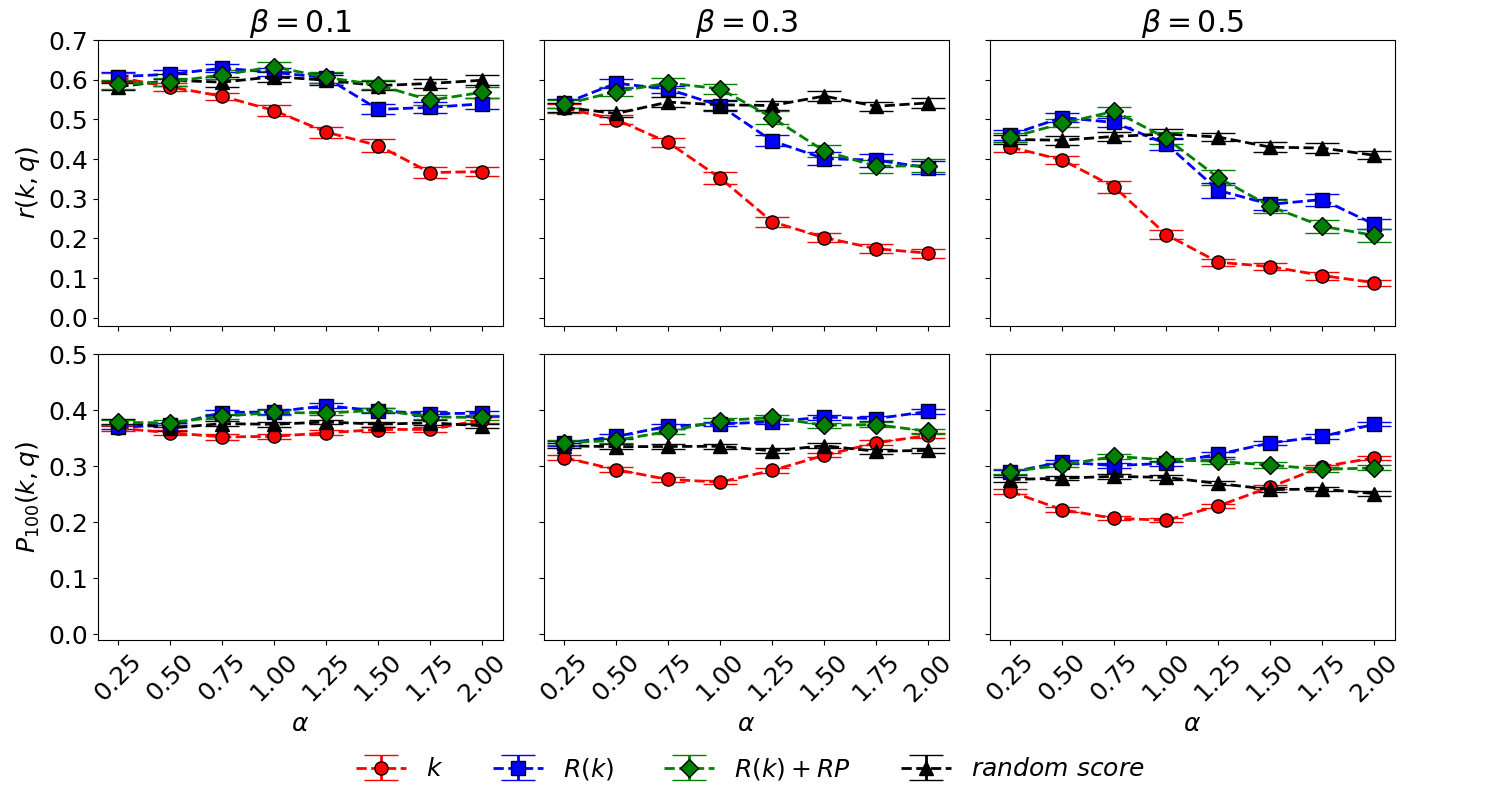}
	\caption{Results for $m=3$. Quality promotion as measured by $r(k, q)$ (the Pearson's linear correlation between node indegree $k$ and node quality $q$ -- \emph{top} panels), and $P_{100}(k, q)$ (the precision of node indegree $k$ in identifying the top-$100$ nodes by quality $q$ -- \emph{bottom} panels): comparison between indegree-generated (circles), $R(k)$-generated (squares), $R(k)+RP$-generated (green rhombuses), and random-generated networks (triangles). The three columns correspond, from left to right, to $\beta=0.1, 0.3, 0.5$, respectively. Results are averaged over $100$ realizations; the error bars represent the standard error of the mean. }
\end{figure*}


\begin{figure*}[h]
\centering
\includegraphics[scale=0.4]{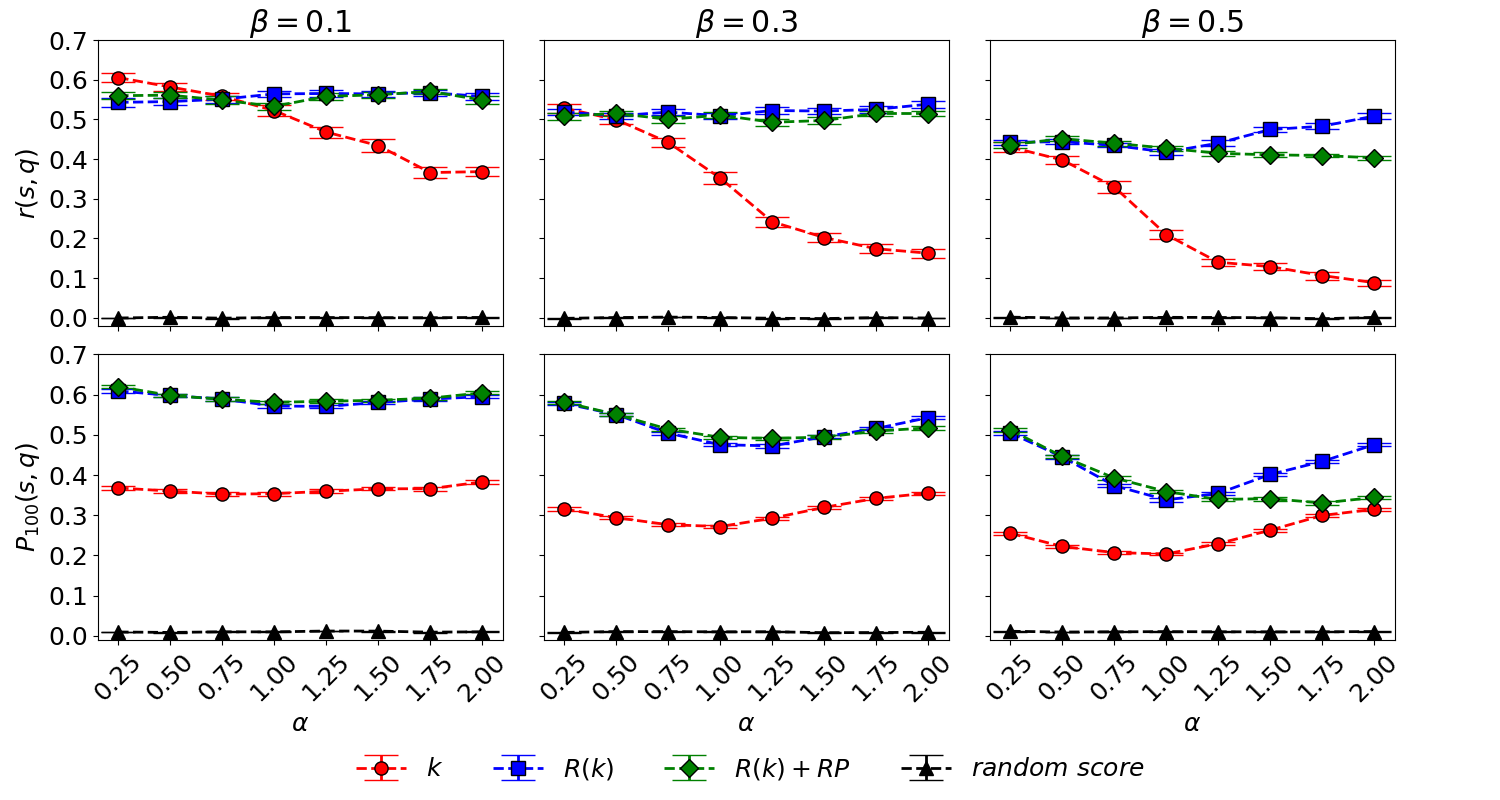}
	\caption{Results for $m=3$. Quality detection as measured by $r(s, q)$ (the Pearson's linear correlation between node score $s$ and node quality $q$ -- \emph{top} panels), and $P_{100}(s, q)$ (the precision of  node score $s$ in identifying the top-$100$ nodes by quality $q$ -- \emph{bottom} panels): comparison between indegree-generated ($s=k$, circles), $R(k)$-generated ($s=R(k)$, $R(k)+RP$-generated (green rhombuses), squares), and random-generated networks ($s=\rho$, triangles). The three columns correspond, from left to right, to $\beta=0.1, 0.3, 0.5$, respectively. The dots represent averages over $100$ realizations; the error bars represent the standard error of the mean.}
\end{figure*}

\begin{figure*}[h]
\centering
\includegraphics[scale=0.4]{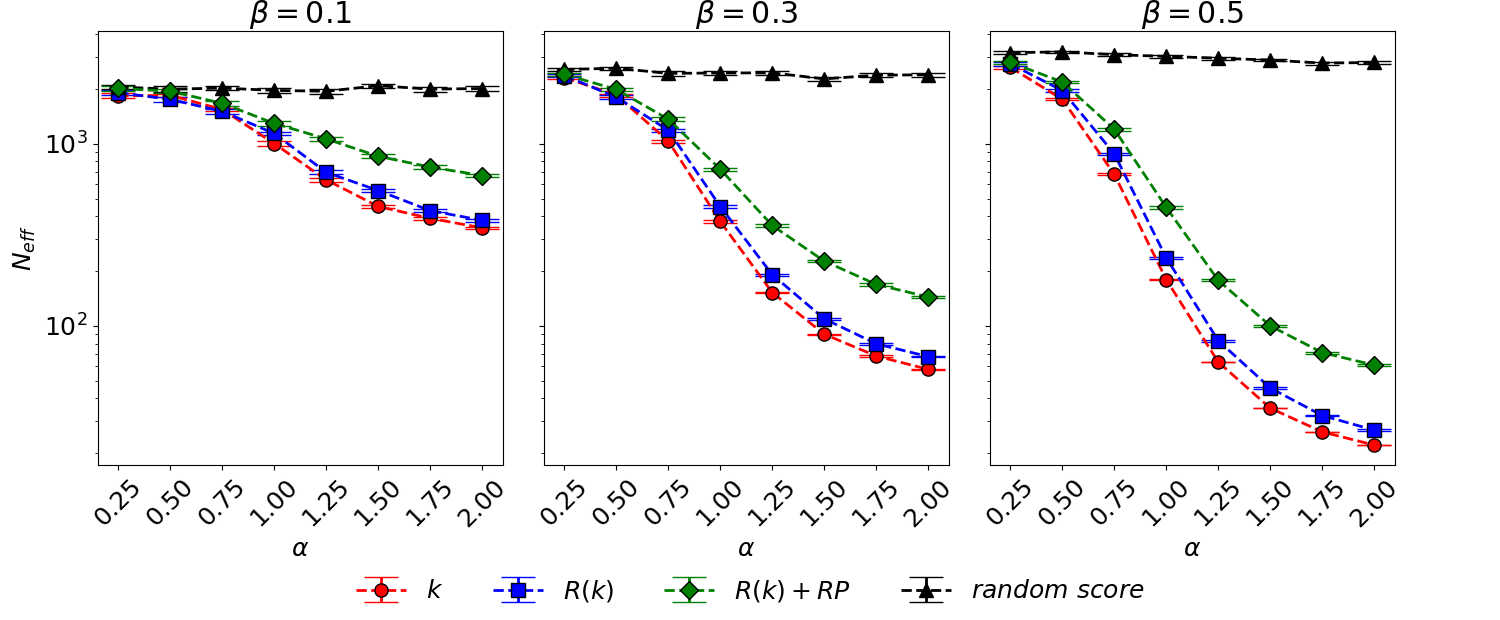}
	\caption{Results for $m=3$. Diversity as measured by the effective number of nodes $N_{eff}$ (the larger, the more egalitarian the indegree distribution): comparison between indegree-generated (circles), $R(k)$-generated (squares), $R(k)+RP$-generated (green rhombuses), and random-generated (triangles) networks. The three columns correspond, from left to right, to $\beta=0.1,0.3,0.5$, respectively. The dots represent averages over $100$ realizations; the error bars represent the standard error of the mean.}
\end{figure*}


\begin{figure*}[h]
\centering
\includegraphics[scale=0.4]{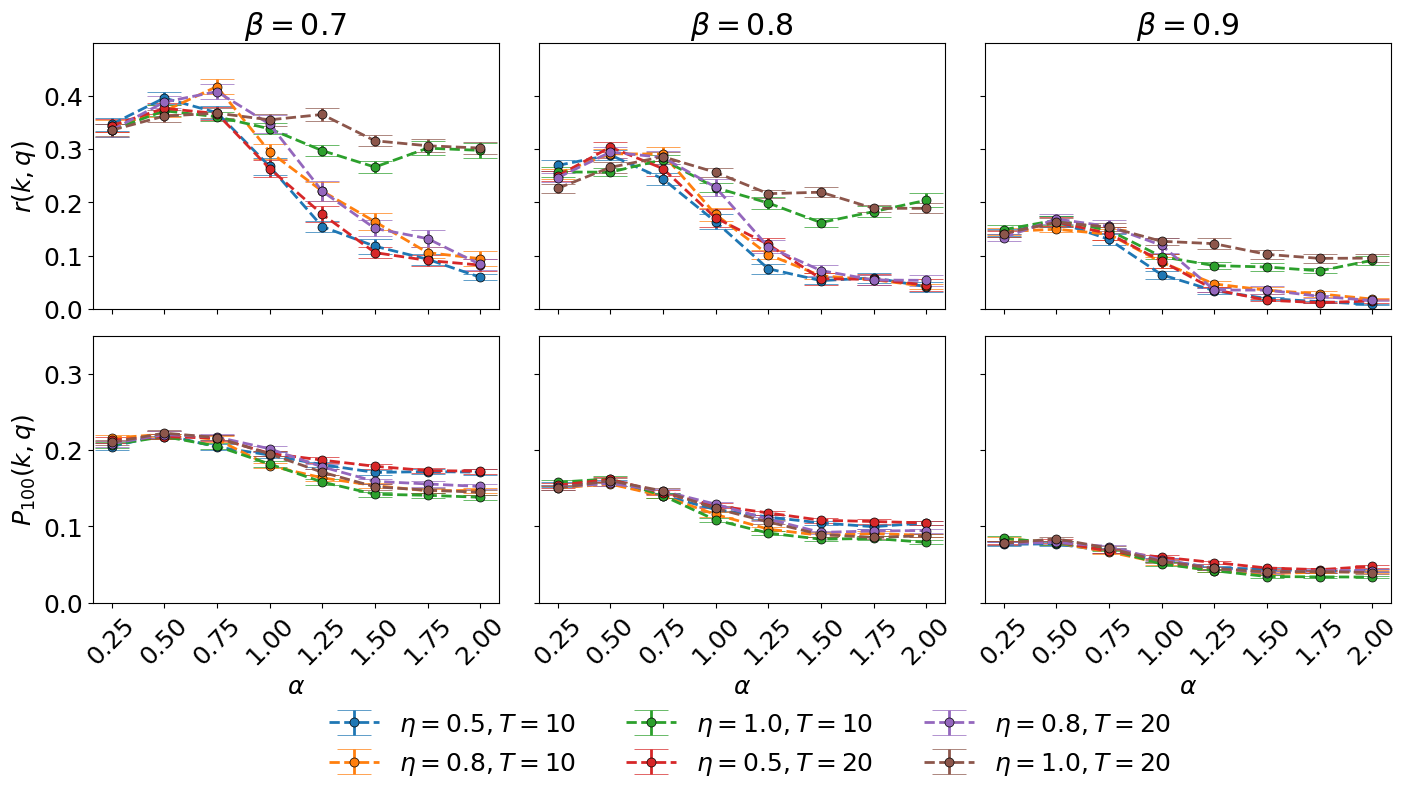}
	\caption{Results for $m=3$. Quality promotion as measured by $r(k, q)$ (the Pearson's linear correlation between node indegree $k$ and node quality $q$ -- \emph{top} panels), and $P_{100}(k, q)$ (the precision of node indegree $k$ in identifying the top-$100$ nodes by quality $q$ -- \emph{bottom} panels): comparison between $R(k)+RP$-generated networks with different parameters $\eta=P/T=5/10, 8/10, 10/10, 10/20, 16/20, 20/20$. The three columns correspond, from left to right, to $\beta=0.7,0.8,0.9$, respectively. The dots represent averages over $100$ realizations; the error bars represent the standard error of the mean.}
\end{figure*}

\begin{figure*}[h]
\centering
\includegraphics[scale=0.4]{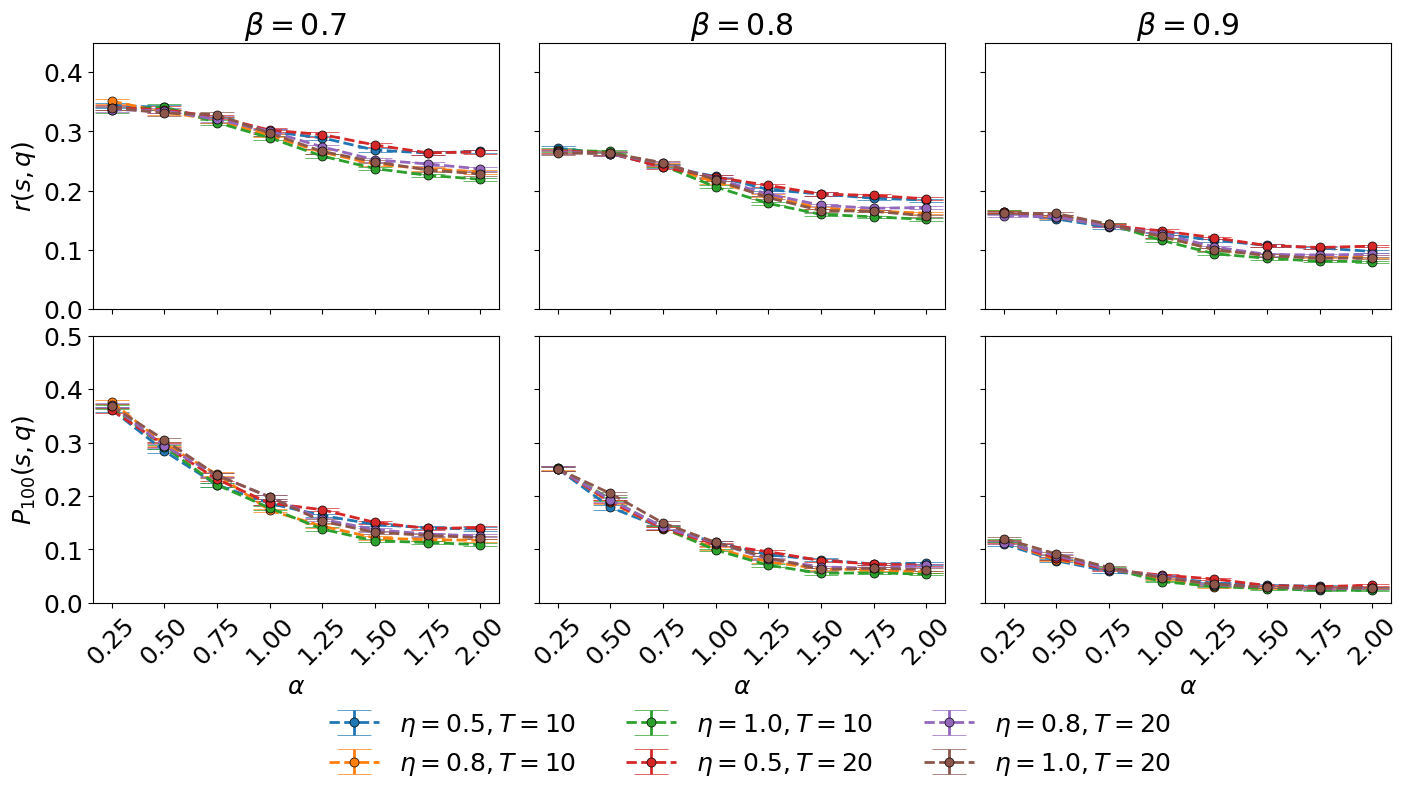}
	\caption{Results for $m=3$. Quality detection as measured by $r(s, q)$ (the Pearson's linear correlation between node score $s$ and node quality $q$ -- \emph{top} panels), and $P_{100}(s, q)$ (the precision of  node score $s$ in identifying the top-$100$ nodes by quality $q$ -- \emph{bottom} panels): comparison between $R(k)+RP$-generated networks with different parameters $\eta=P/T=5/10, 8/10, 10/10, 10/20, 16/20, 20/20$. The three columns correspond, from left to right, to $\beta=0.7,0.8,0.9$, respectively. The dots represent averages over $100$ realizations; the error bars represent the standard error of the mean.}
\end{figure*}


\begin{figure*}[h]
\centering
\includegraphics[scale=0.4]{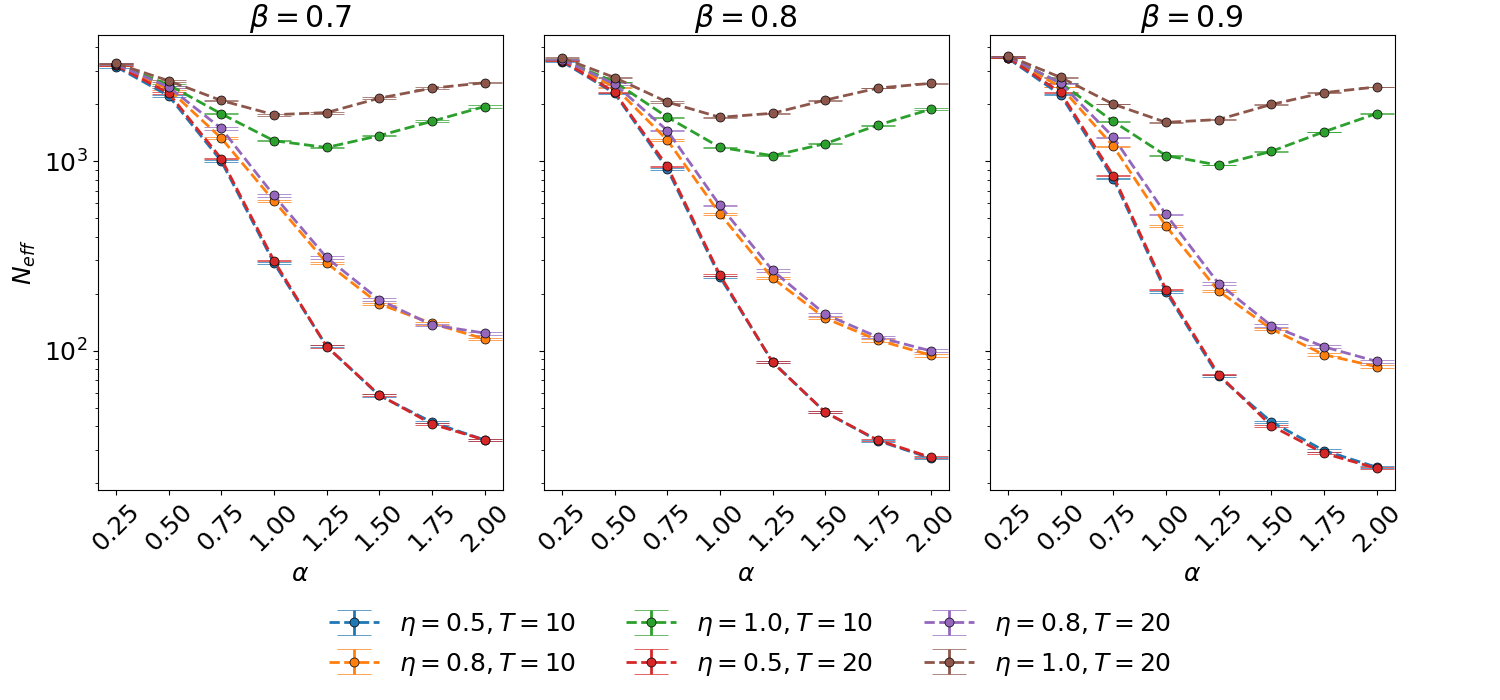}
	\caption{Results for $m=3$. Diversity as measured by the effective number of nodes $N_{eff}$ (the larger, the more egalitarian the indegree distribution): comparison between $R(k)+RP$-generated networks with different parameters $\eta=P/T=5/10, 8/10, 10/10, 10/20, 16/20, 20/20$. The three columns correspond, from left to right, to $\beta=0.7,0.8,0.9$, respectively. The dots represent averages over $100$ realizations; the error bars represent the standard error of the mean.}
\end{figure*}


\begin{figure*}[h]
\centering
\includegraphics[scale=0.3]{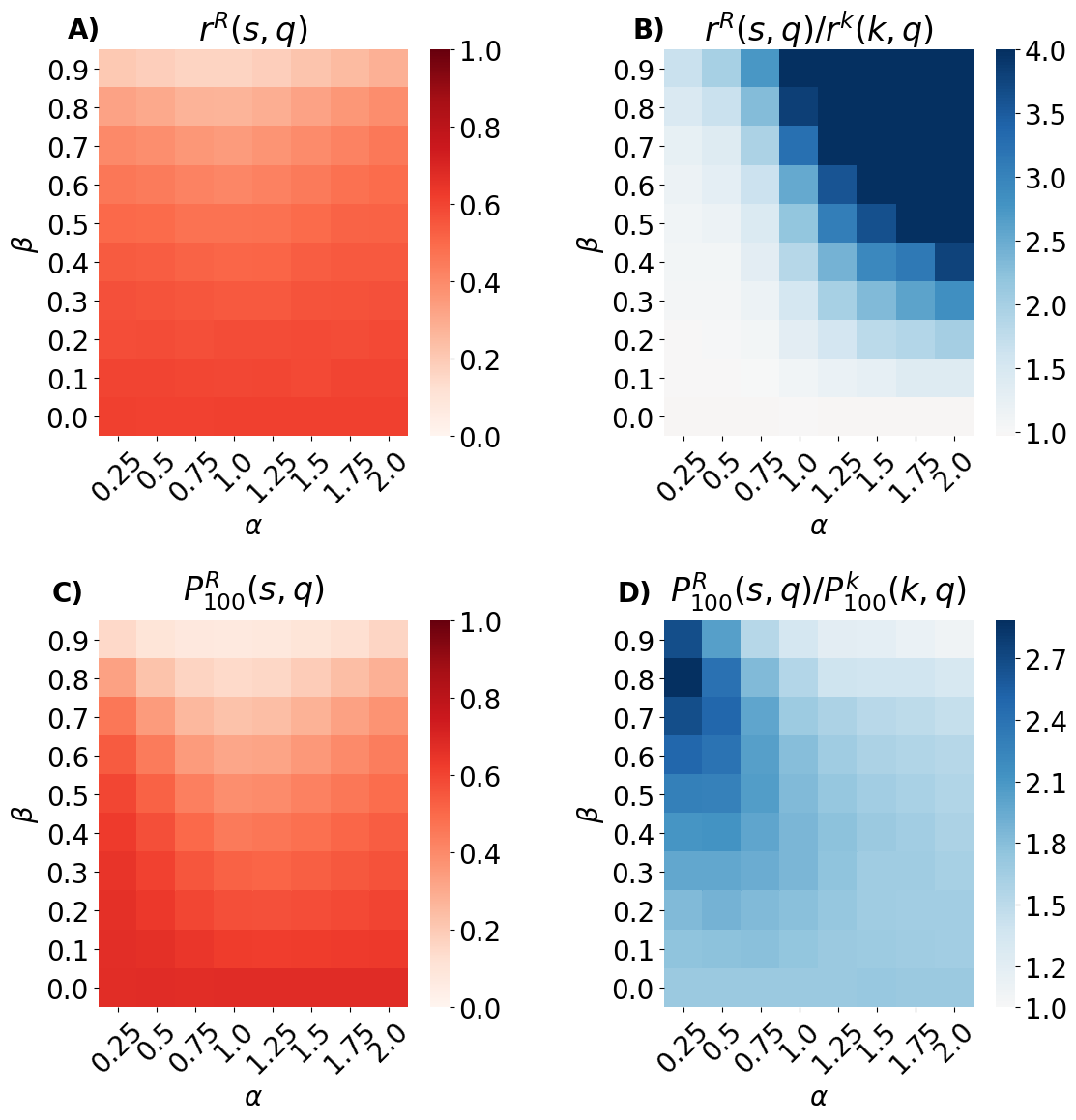}
	\caption{Results for $m=6$. Quality detection as measured by $r(s, q)$ (the Pearson's linear correlation between 
node score $s$ and node quality $q$ -- \emph{top} panels), and $P_{100}(s, q)$ (the precision of node score $s$ in identifying the top-100 nodes by quality $q$ -- \emph{bottom} panels): comparison between popularity-generated and $R(k)$-generated networks. (A) $r^R(s, q)$ for $R(k)$-generated networks, as a function of the model parameters $\alpha$ (exploration cost) and $\beta$ (popularity bias). (B) Ratio $r^{R}(s,q)/r^{k}(k,q)$ as a function of the model parameters. (C) $P_{100}(s,q)$ for $R(k)$-generated networks, as a function of the model parameters. (D) Ratio $P_{100}^{R}(s,q)/P_{100}^{k}(k,q)$ as a function of the model parameters. Results are averaged over $500$ realizations.}
\end{figure*}

\begin{figure*}[h]
\centering
\includegraphics[scale=0.3]{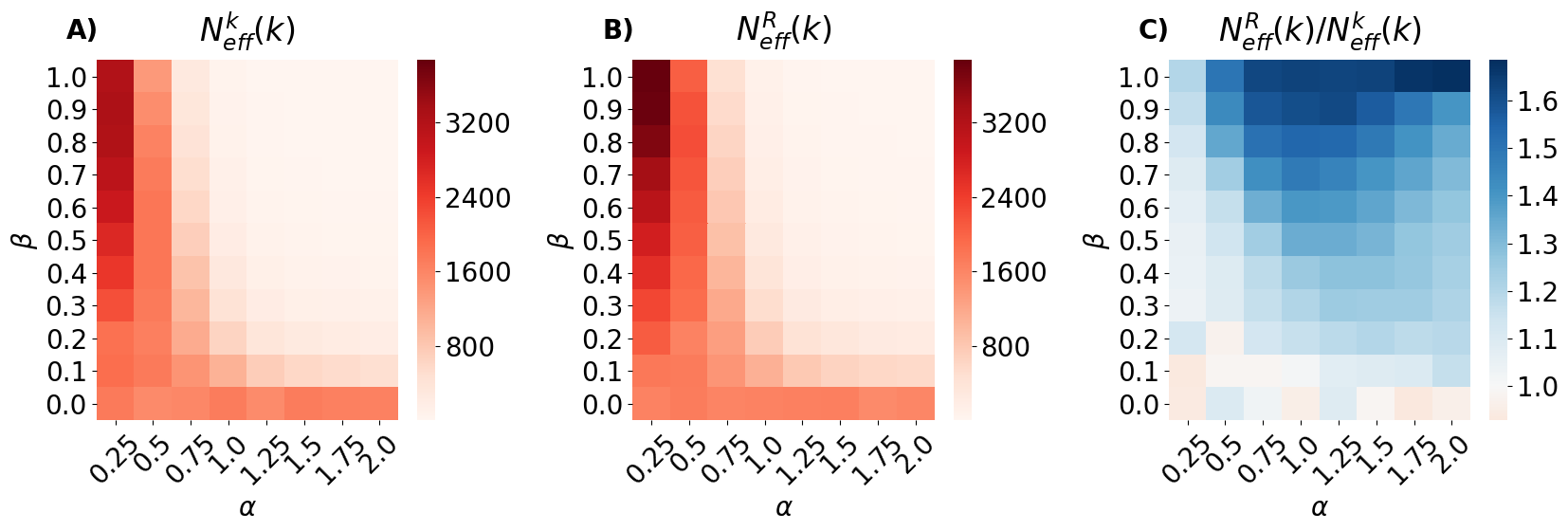}
	\caption{Results for $m=6$. Diversity as measured by the effective number of nodes: comparison between popularity-generated and $R(k)$-generated networks. (A-B) The effective number of nodes $N_{eff}(k)$ for indegree-generated ($N_{eff}^k(k)$, panel A) and $R(k)$-generated ($N_{eff}^R(k)$, panel B) networks, as a function of the model parameters. (C) Ratio $N_{eff}^R(k)/N_{eff}^k(k)$ as a function of the model parameters. Results are averaged over $500$ realizations.}
\end{figure*}


\begin{figure*}[h]
\centering
\includegraphics[scale=0.4]{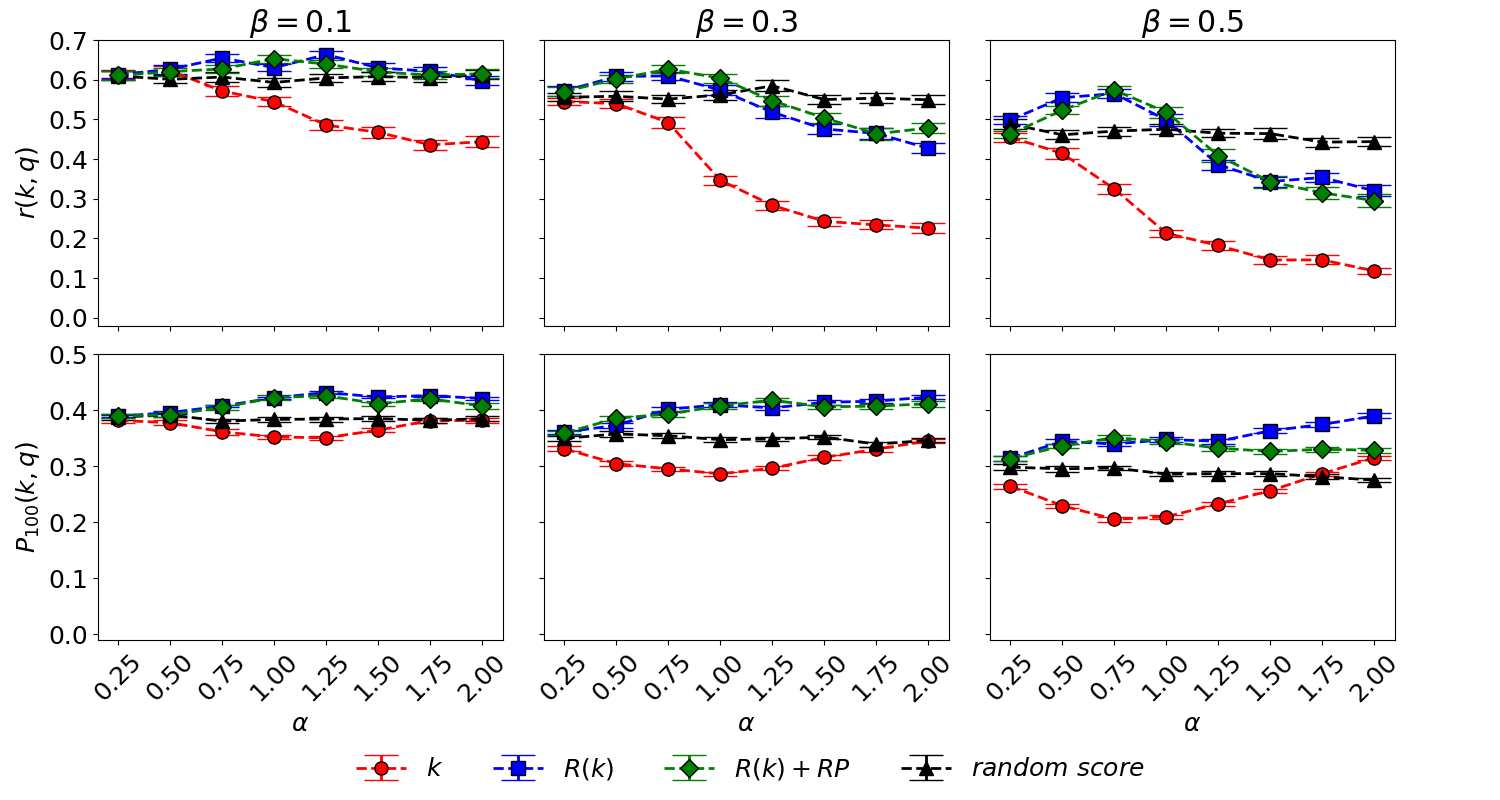}
	\caption{Results for $m=6$. Quality promotion as measured by $r(k, q)$ (the Pearson's linear correlation between node indegree $k$ and node quality $q$ -- \emph{top} panels), and $P_{100}(k, q)$ (the precision of node indegree $k$ in identifying the top-$100$ nodes by quality $q$ -- \emph{bottom} panels): comparison between indegree-generated (circles), $R(k)$-generated (squares), $R(k)+RP$-generated (green rhombuses), and random-generated networks (triangles). The three columns correspond, from left to right, to $\beta=0.1, 0.3, 0.5$, respectively. Results are averaged over $100$ realizations; the error bars represent the standard error of the mean. }
\end{figure*}

\begin{figure*}[h]
\centering
\includegraphics[scale=0.4]{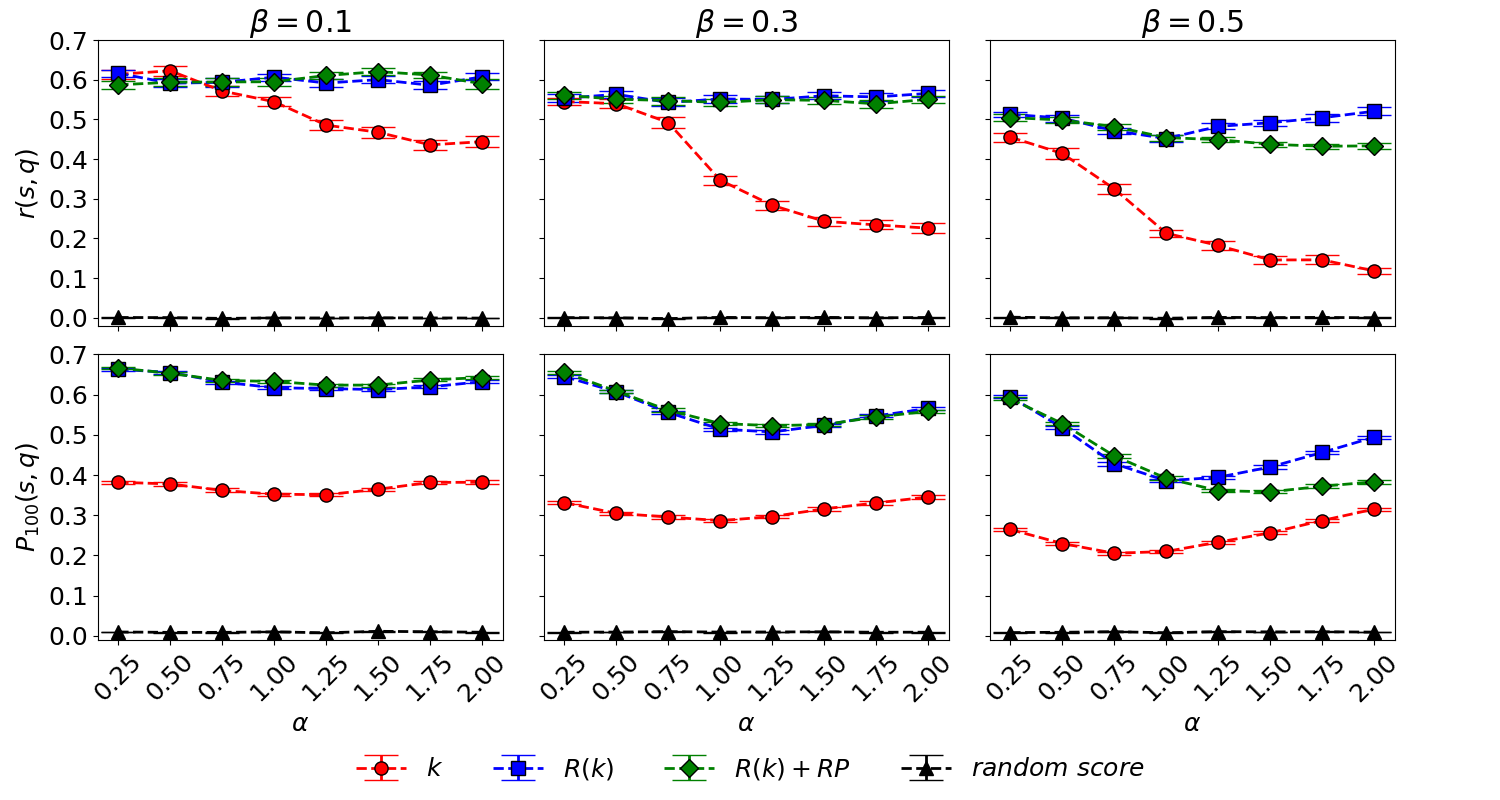}
	\caption{Results for $m=6$. Quality detection as measured by $r(s, q)$ (the Pearson's linear correlation between node score $s$ and node quality $q$ -- \emph{top} panels), and $P_{100}(s, q)$ (the precision of  node score $s$ in identifying the top-$100$ nodes by quality $q$ -- \emph{bottom} panels): comparison between indegree-generated ($s=k$, circles), $R(k)$-generated ($s=R(k)$, squares), $R(k)+RP$-generated (green rhombuses), and random-generated networks ($s=\rho$, triangles). The three columns correspond, from left to right, to $\beta=0.1, 0.3, 0.5$, respectively. The dots represent averages over $100$ realizations; the error bars represent the standard error of the mean. }
\end{figure*}


\begin{figure*}[h]
\centering
\includegraphics[scale=0.4]{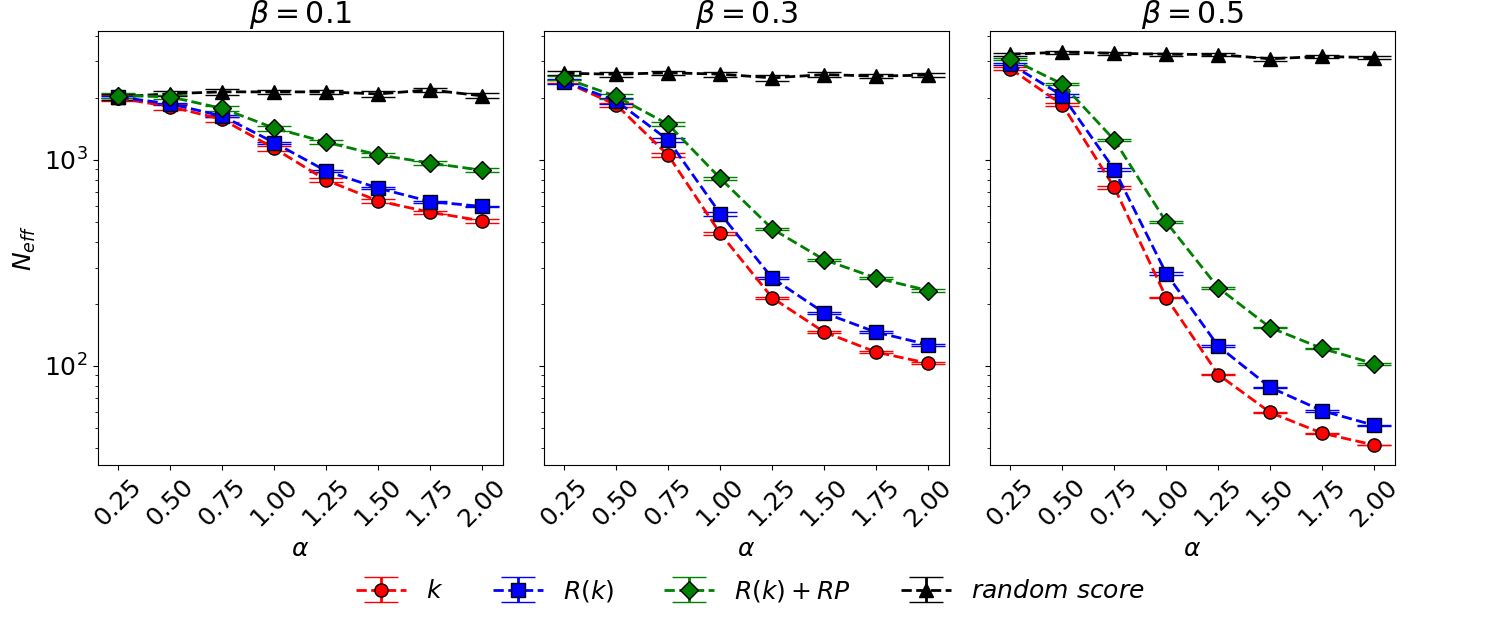}
	\caption{Results for $m=6$. Diversity as measured by the effective number of nodes $N_{eff}$ (the larger, the more egalitarian the indegree distribution): comparison between indegree-generated (circles), $R(k)$-generated (squares), $R(k)+RP$-generated (green rhombuses), and random-generated (triangles) networks. The three columns correspond, from left to right, to $\beta=0.1,0.3,0.5$, respectively. The dots represent averages over $100$ realizations; the error bars represent the standard error of the mean.}
\end{figure*}

\begin{figure*}[h]
\centering
\includegraphics[scale=0.4]{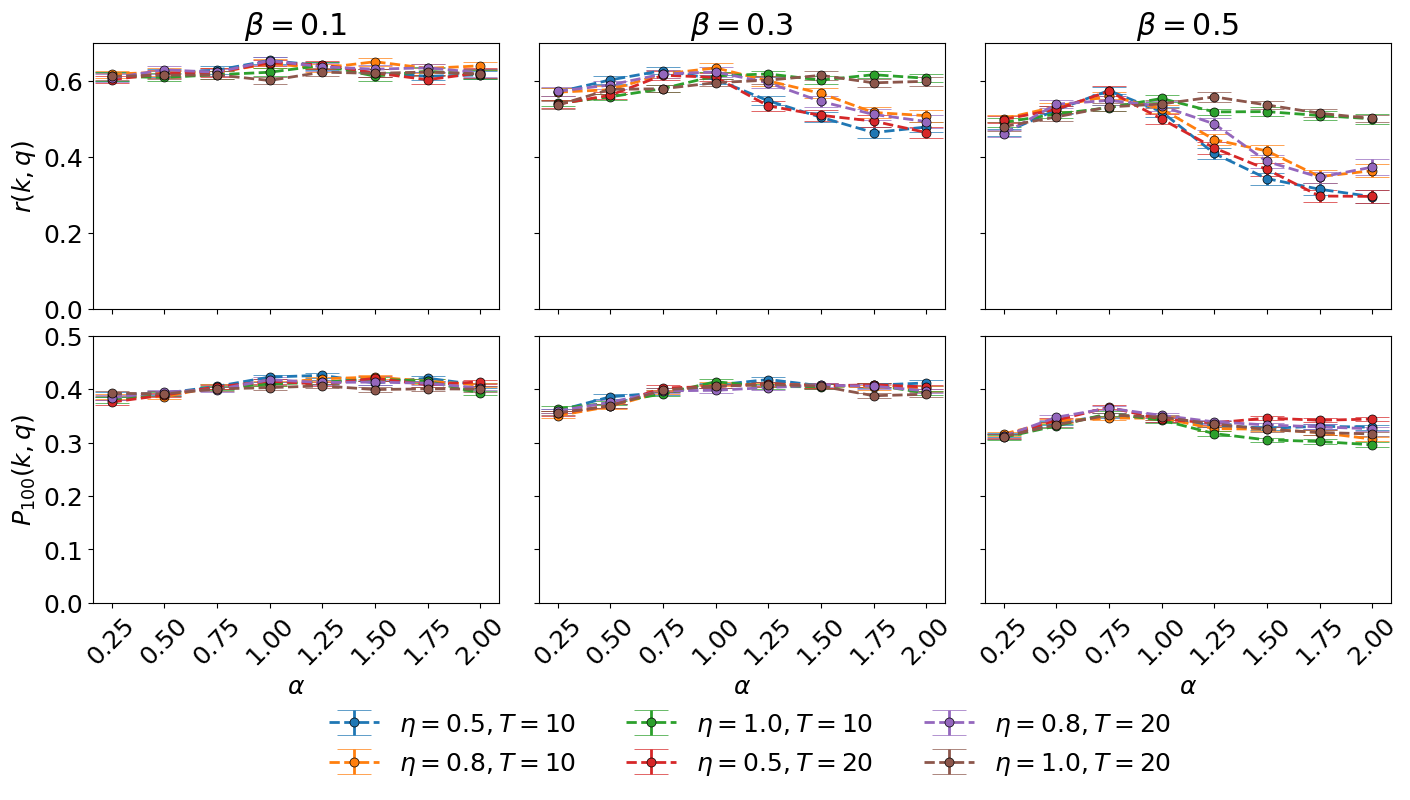}
	\caption{Results for $m=6$. Quality promotion as measured by $r(k, q)$ (the Pearson's linear correlation between node indegree $k$ and node quality $q$ -- \emph{top} panels), and $P_{100}(k, q)$ (the precision of node indegree $k$ in identifying the top-$100$ nodes by quality $q$ -- \emph{bottom} panels): comparison between $R(k)+RP$-generated networks with different parameters $\eta=P/T=5/10, 8/10, 10/10, 10/20, 16/20, 20/20$. The three columns correspond, from left to right, to $\beta=0.1,0.3,0.5$, respectively. The dots represent averages over $100$ realizations; the error bars represent the standard error of the mean.}
\end{figure*}


\begin{figure*}[h]
\centering
\includegraphics[scale=0.4]{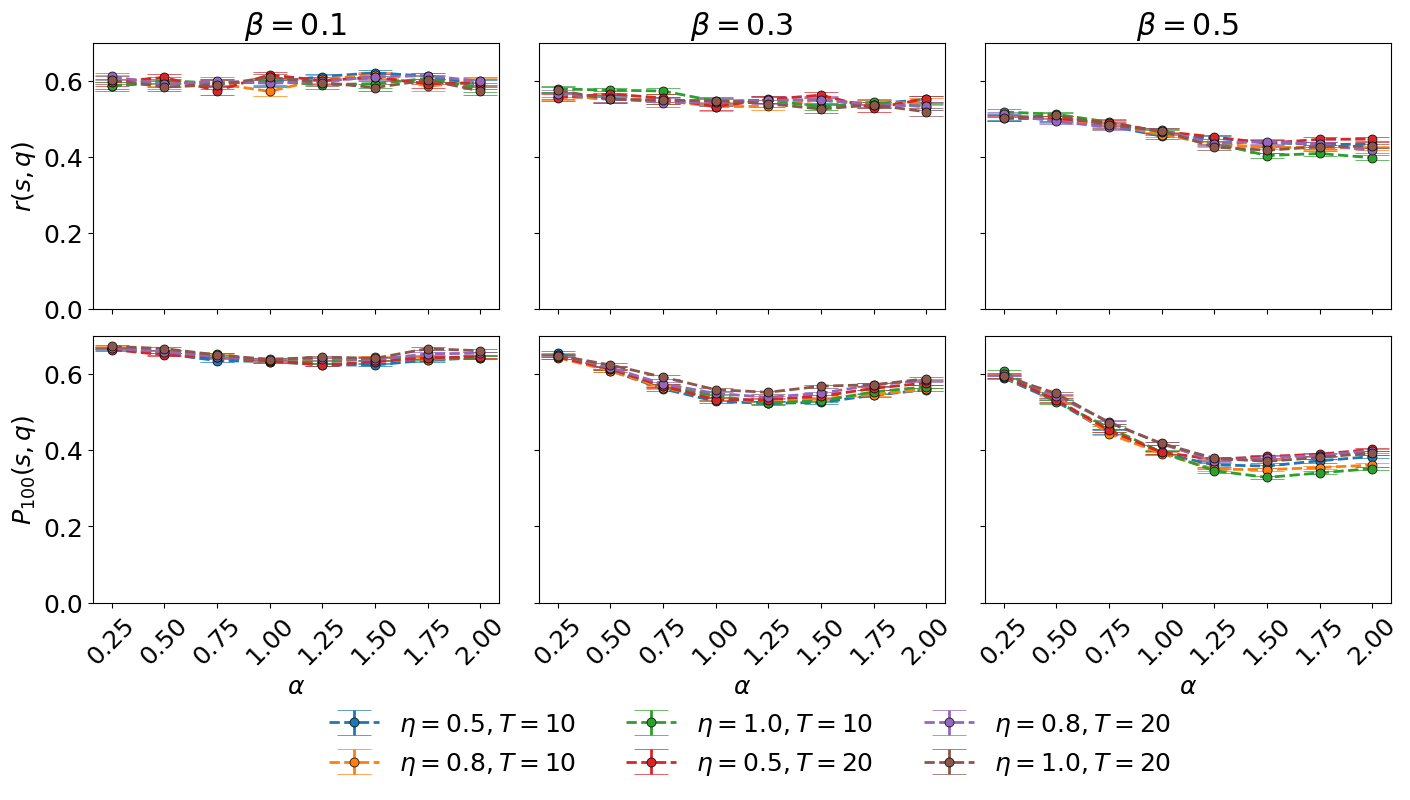}
	\caption{Results for $m=6$. Quality detection as measured by $r(s, q)$ (the Pearson's linear correlation between node score $s$ and node quality $q$ -- \emph{top} panels), and $P_{100}(s, q)$ (the precision of  node score $s$ in identifying the top-$100$ nodes by quality $q$ -- \emph{bottom} panels): comparison between $R(k)+RP$-generated networks with different parameters $\eta=P/T=5/10, 8/10, 10/10, 10/20, 16/20, 20/20$. The three columns correspond, from left to right, to $\beta=0.1,0.3,0.5$, respectively. The dots represent averages over $100$ realizations; the error bars represent the standard error of the mean.}
\end{figure*}

\begin{figure*}[h]
\centering
\includegraphics[scale=0.4]{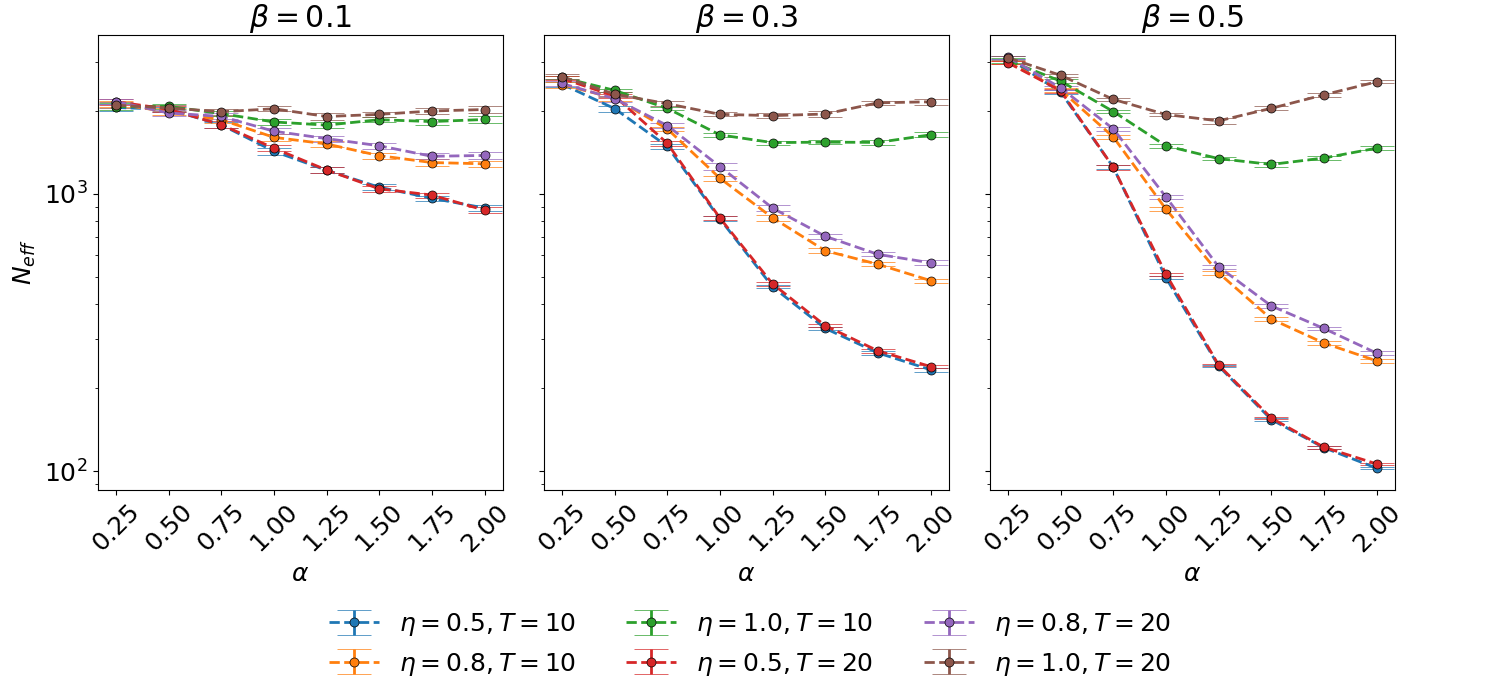}
	\caption{Results for $m=6$. Diversity as measured by the effective number of nodes $N_{eff}$ (the larger, the more egalitarian the indegree distribution): comparison between $R(k)+RP$-generated networks with different parameters $\eta=P/T=5/10, 8/10, 10/10, 10/20, 16/20, 20/20$. The three columns correspond, from left to right, to $\beta=0.1,0.3,0.5$, respectively. The dots represent averages over $100$ realizations; the error bars represent the standard error of the mean.}
\end{figure*}


\begin{figure*}[h]
\centering
\includegraphics[scale=0.4]{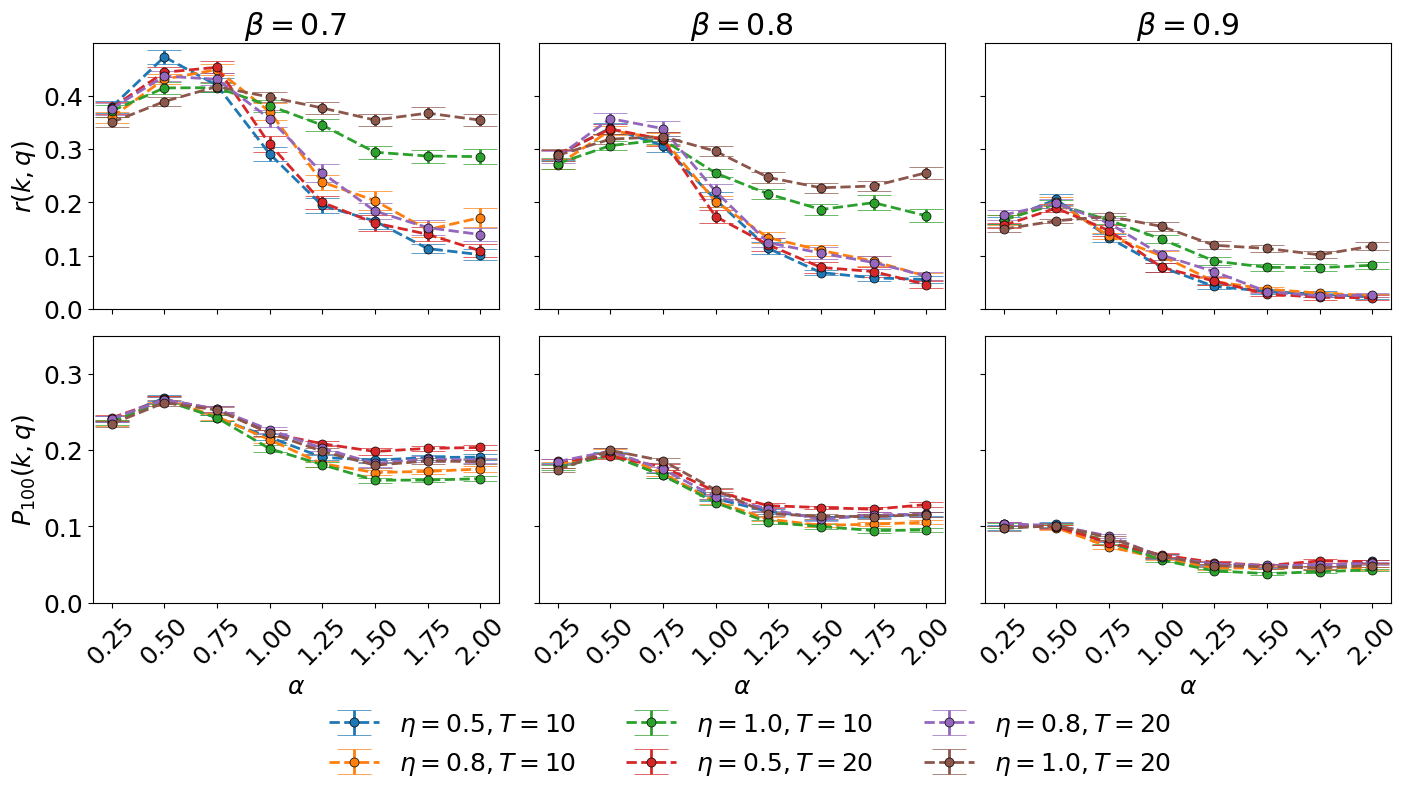}
	\caption{Results for $m=6$. Quality promotion as measured by $r(k, q)$ (the Pearson's linear correlation between node indegree $k$ and node quality $q$ -- \emph{top} panels), and $P_{100}(k, q)$ (the precision of node indegree $k$ in identifying the top-$100$ nodes by quality $q$ -- \emph{bottom} panels): comparison between $R(k)+RP$-generated networks with different parameters $\eta=P/T=5/10, 8/10, 10/10, 10/20, 16/20, 20/20$. The three columns correspond, from left to right, to $\beta=0.7,0.8,0.9$, respectively. The dots represent averages over $100$ realizations; the error bars represent the standard error of the mean.}
\end{figure*}


\begin{figure*}[h]
\centering
\includegraphics[scale=0.4]{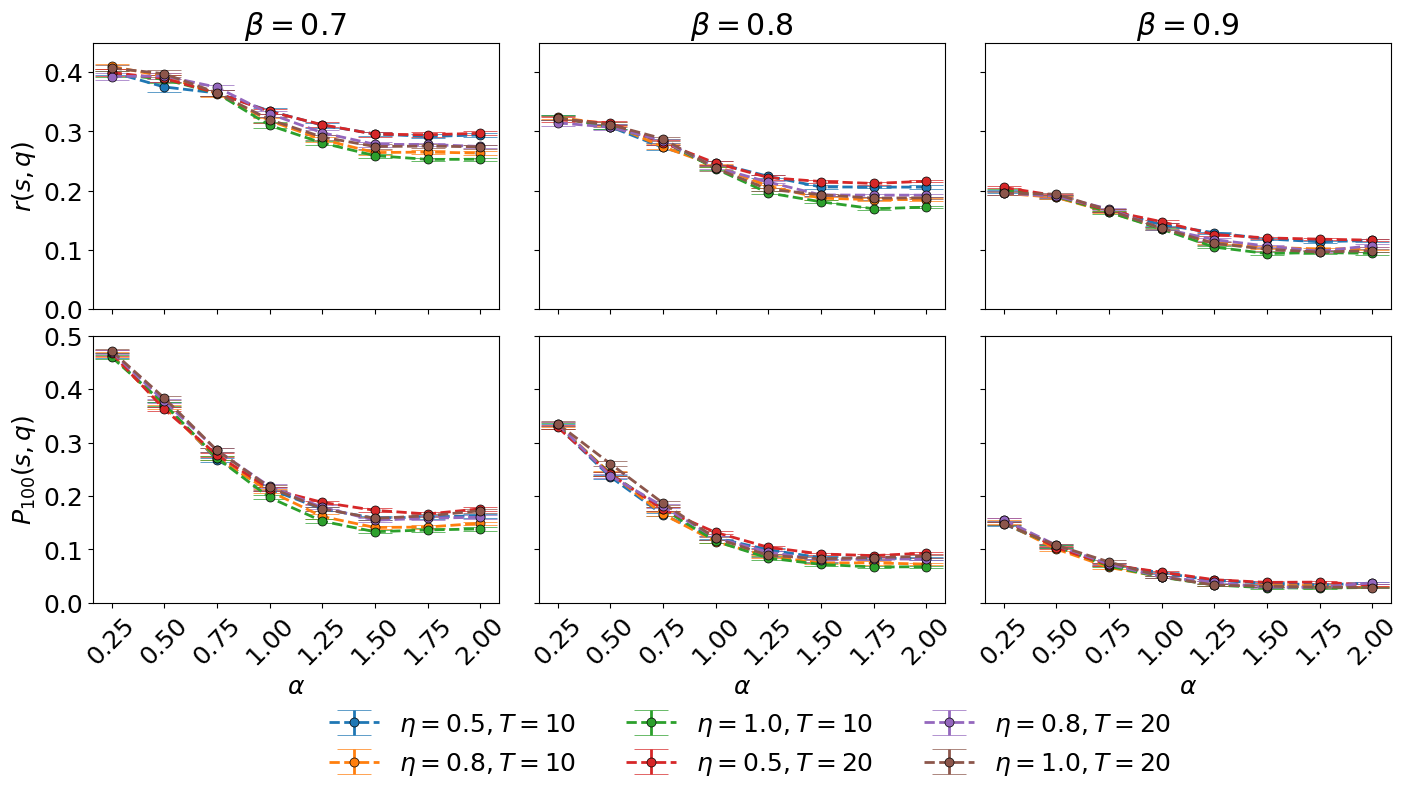}
	\caption{Results for $m=6$. Quality detection as measured by $r(s, q)$ (the Pearson's linear correlation between node score $s$ and node quality $q$ -- \emph{top} panels), and $P_{100}(s, q)$ (the precision of  node score $s$ in identifying the top-$100$ nodes by quality $q$ -- \emph{bottom} panels): comparison between $R(k)+RP$-generated networks with different parameters $\eta=P/T=5/10, 8/10, 10/10, 10/20, 16/20, 20/20$. The three columns correspond, from left to right, to $\beta=0.7,0.8,0.9$, respectively. The dots represent averages over $100$ realizations; the error bars represent the standard error of the mean.}
\end{figure*}

\begin{figure*}[h]
\centering
\includegraphics[scale=0.4]{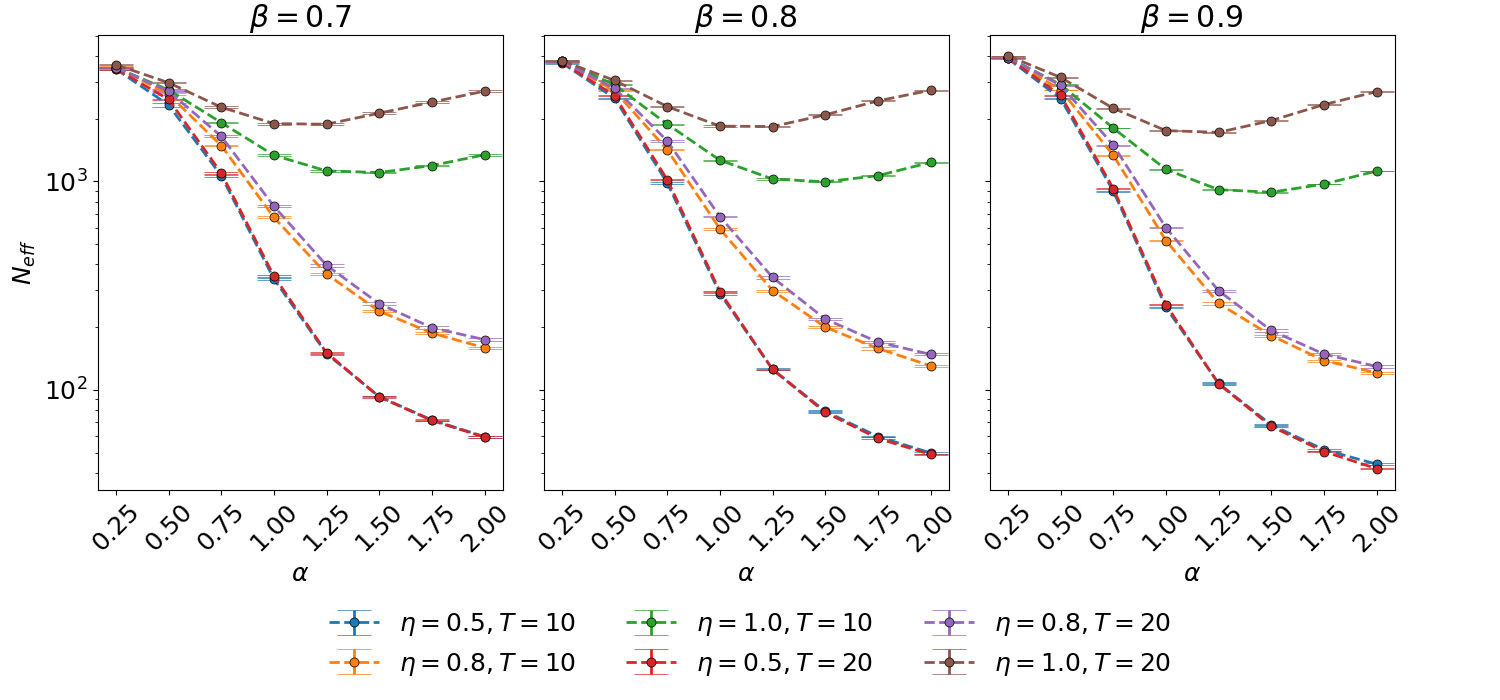}
	\caption{Results for $m=6$. Diversity as measured by the effective number of nodes $N_{eff}$ (the larger, the more egalitarian the indegree distribution): comparison between $R(k)+RP$-generated networks with different parameters $\eta=P/T=5/10, 8/10, 10/10, 10/20, 16/20, 20/20$. The three columns correspond, from left to right, to $\beta=0.7,0.8,0.9$, respectively. The dots represent averages over $100$ realizations; the error bars represent the standard error of the mean.}
\end{figure*}


\begin{figure*}[h]
\centering
\includegraphics[scale=0.4]{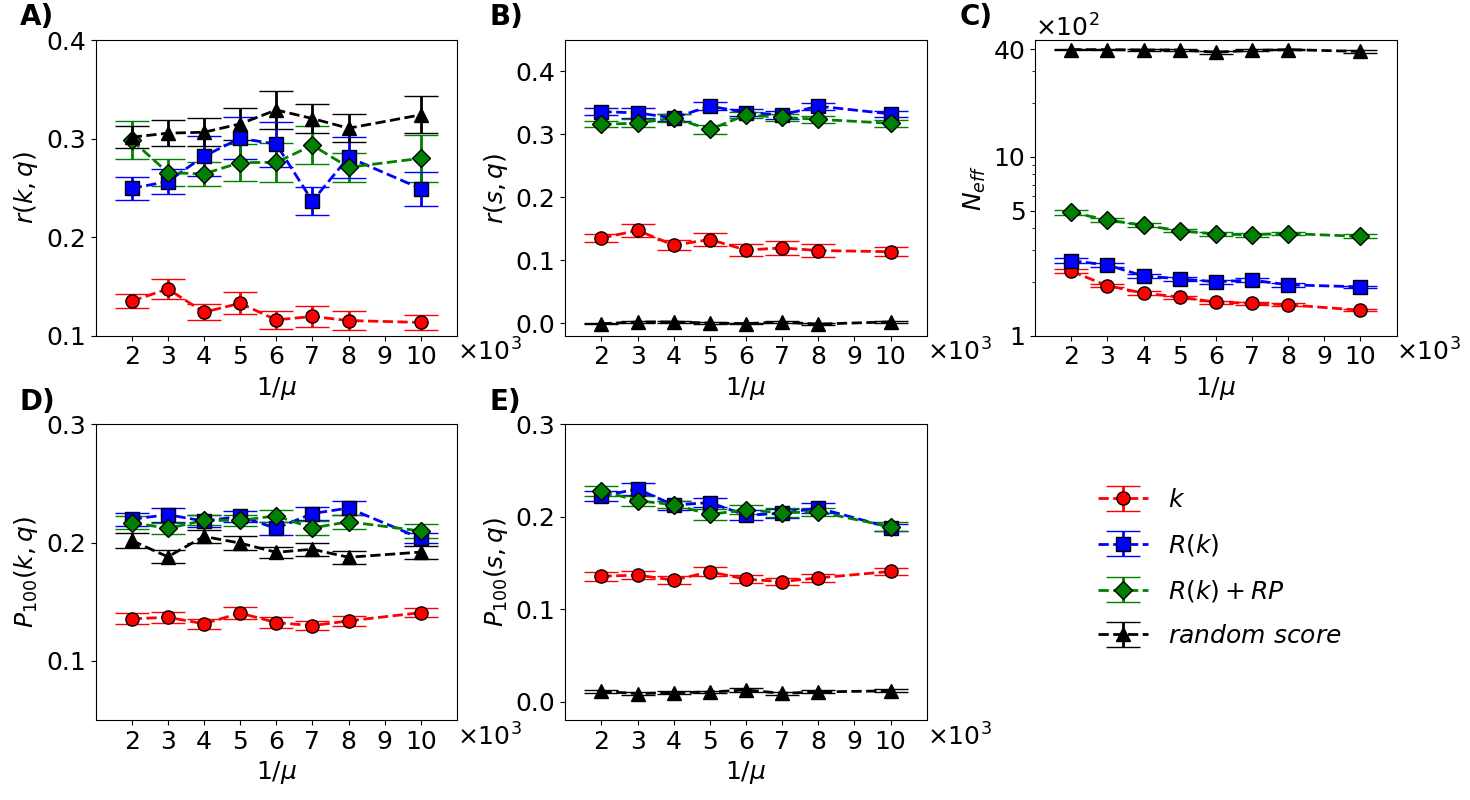}
	\caption{Results for model including node removal, $m=6$. The impact of node removal on quality promotion, quality detection, popularity diversity. For $\alpha=1.0$ and $\beta= 0.7$, we show five network properties as a function of the inverse $A^{*}=\mu^{-1}$ of the removal probability: (A) the Pearson's correlation between node indegree and node quality; (B) the Pearson's correlation between node score and node quality; (C) the effective number of nodes, $N_{eff}$; (D) the indegree's precision in identifying the top-$100$ nodes by quality; (E) the score's precision in identifying the top-$100$ nodes by quality. Different lines correspond to the networks generated with different algorithms. The dots represent averages over $50$ realizations; the error bars represent the standard error of the mean.}
\end{figure*}


\begin{figure*}][h]
\centering
\includegraphics[scale=0.4]{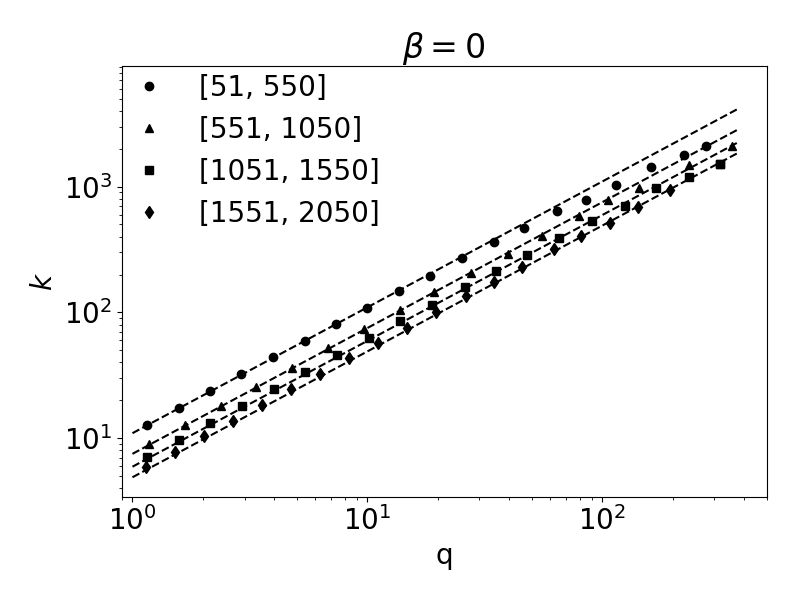}
	\caption{Results for $m=6$. The relation between node indegree and quality in networks of $N=10,000$ nodes generated with $\beta=0$. To factor out the dependence of node indegree on node age, we group together nodes of similar age -- we represent here the results for four such age groups. The analytic lines well match the results of numerical simulations for sufficiently small $q$ values. }
\end{figure*}

\begin{figure*}[h]
\centering
\includegraphics[scale=0.3]{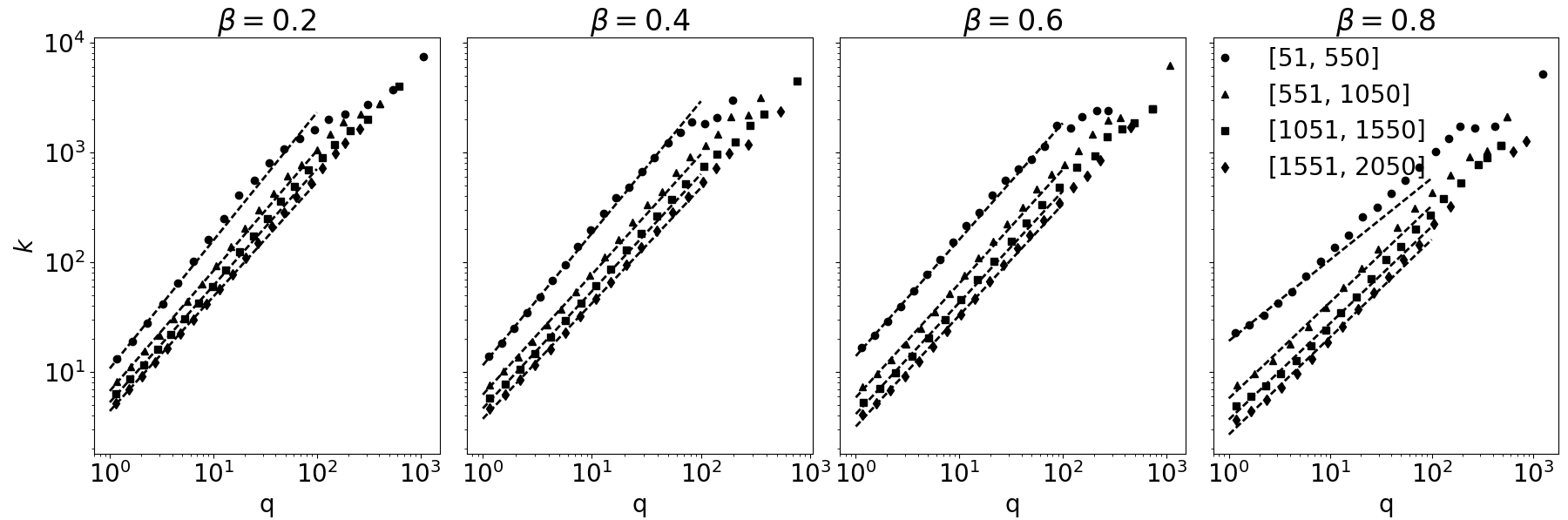}
	\caption{Results for $m=6$. The relation between node indegree and quality in networks of $N=10,000$ nodes generated with $\beta=0,2,0.4,0.6,0.8$. To factor out the dependence of node indegree on node age, we group together nodes of similar age -- we represent here the results for four such age groups. The results of a power-law fit (see Table S1) well match the numerical results for small $q$ values. }
\end{figure*}

\clearpage

\begin{center}
\begin{table}[!h]
  \centering
    \setlength{\tabcolsep}{2mm}{\begin{tabular}{|c|c|c|c|c|c|c|c|c|}
	\hline
    \multirow{2}*{Age group} & \multicolumn{2}{c|}{$\beta=0.2$} & \multicolumn{2}{c|}{$\beta=0.4$} & \multicolumn{2}{c|}{$\beta=0.6$} & \multicolumn{2}{c|}{$\beta=0.8$} \\ \cline{2-9}
     & A & $\delta$ & A & $\delta$ & A & $\delta$ & A & $\delta$ \\
    \hline
    $[51, 550]$ & 10.80 & 1.16 & 11.54 & 1.20 & 14.02 & 1.06 & 19.23 & 0.74 \\
    $[551, 1050]$ & 6.73 & 1.09 & 6.24 & 1.09 & 5.89 & 1.04 & 5.78 & 0.88 \\
    $[1051, 1550]$ & 5.31 & 1.06 & 4.68 & 1.07 & 4.16 & 1.01 & 3.71 & 0.88 \\
    $[1551, 2050]$ & 4.42 & 1.04 & 3.76 & 1.05 & 3.20 & 1.01 & 2.71 & 0.89 \\
    \hline
    \end{tabular}}%
  \caption{Results for $m=6$. Results of the fit of the relation between $k$ and $q$ (Fig. S17) using $\overline{k}=A_i q^{\delta_i}$; To account for temporal effects, we split the nodes into groups of $500$ nodes based on their age (e.g., $[551,1050]$). For each age group, the fit is performed as a least-squares linear regression of the relation $\log{k}=\log{A_i}+\delta_i\,\log{q}$, for $q\leq 100$. We find that the fit results well match the numerical results (see Fig. S17), and the exponent $\delta_i$ depends weakly on the age group $i$.}
  \label{tab:addlabel}%
\end{table}%
\end{center}
\end{document}